\documentclass[twocolumn]{jpsj2} 
\usepackage{subfigure}
\title{Green's Function of Fully Anharmonic Lattice Vibration}

\author{Masateru Takechi and Kazuo Ueda}
\inst{Institute for Solid State Physics, University of Tokyo, Kashiwanoha, Kashiwa, Chiba 277-8581, Japan}

\date{\today}

\abst{Motivated by the discovery of superconductivity in $\beta$-pyrochlore oxides, we study property of rattling motion
coupled with conduction electrons.
We derive the general expression of the Green's function of fully anharmonic lattice vibration
within the accuracy of the second order perturbation of electron-ion interaction
by introducing self-energy, vertex-correction, and normalization factor
for each transition. Using the expression, we discuss the characteristic properties of the spectral function
in the entire range from weakly anharmonic potential to
double-well case, and calculate NMR relaxation rate due to the two phonon Raman process.}

\kword{rattling, anharmonicity}

\begin{document}
\setlength{\textwidth}{504pt}
\setlength{\columnsep}{14pt}
\hoffset-23.5pt
\maketitle

\section{Introduction}

Recently discovered $\beta$-pyrochlore oxides, ${\rm KOs_{2}O_{6}}$, ${\rm RbOs_{2}O_{6}}$, and ${\rm CsOs_{2}O_{6}}$,
show unusual properties, which originate from
anharmonic lattice vibration. In these compounds, guest alkaline ion is enclosed in an oversized
cage made of ${\rm Os}$-${\rm O}$ octahedra. 
As the radius of the guest ion gets smaller, it can move around in a larger area subject to weaker restoring force around its center and
consequently shows larger atomic displacement\cite{structure}.
  
Resistivity of ${\rm KOs_{2}O_{6}}$ shows concave-downward temperature dependence\cite{resistivity}. 
In the high temperature limit, electron-ion scattering
dominates the resistivity, which is asymptotically proportional to the ion cross-section. By using WKB approximation\cite{wkb}, one can show that
thermal average of the square of the ion 
displacement also has concave-downward temperature dependence in the high temperature limit when the ion oscillates in an anharmonic potential.
NMR relaxation rate at the K site in ${\rm KOs_{2}O_{6}}$ shows a peculiar temperature dependence, 
which has a low temperature peak and then approaches to a constant 
value as temperature is increased\cite{nmr1}. Dahm and one of the authors explained this phenomena by calculating NMR relaxation rate due to the
two phonon Raman process($1/T_1^RT$) 
introducing a self-consistent quasiharmonic approximation\cite{DU}. They showed that one important aspect of the effect of anharmonicity can be understood as 
the shift of ion spectral weight to higher energy with increasing temperature. 

Now, we turn our attention to the mass-enhancement and superconductivity as interesting phenomena concerning the coupling between
electrons and anharmonic lattice vibration.
Low-temperature electronic specific heat measurements show that effective mass gets progressively large as the radius of the guest ion gets small.
${\rm KOs_{2}O_{6}}$ shows the mass-enhancement about twice of that in ${\rm RbOs_{2}O_{6}}$\cite{mass1,mass2}. This tendency of mass-enhancement
depending on the guest ion radius is one experimental proof of the coupling of electrons with anharmonic lattice vibration.
On the other hand, the Einstein frequency for these compounds is not proportional to the inverse of the square root of the ion mass,
and even gets smaller with decreasing ion mass\cite{Hiroi2005, mass1}. 
That is because the restoring force gets weaker with the smaller ion radius in the oversized cage. Actually the 
band structure calculations confirm this tendency\cite{band1,band2}. 
However, these compounds exhibit progressively higher $T_{\rm c}$ with decreasing Einstein frequency.
This unusual tendency which is not consistent with the well-known theory of superconductivity is another proof of the coupling 
with anharmonic lattice vibration, and indicates that anharmonic lattice vibration can possibly result in higher $T_{\rm c}$.
To the best of the authors' knowledge, the rigorous modification to Eliashberg's theory\cite{Eliashberg1,Eliashberg2}
which makes it possible to treat full effect of anharmonicity
and reveal the relation between anharmonicity and electron-ion coupling does not exist. Then, this problem is left as an open question to study.
As the first step toward this direction, we will study theoretical treatment of anharmonic ionic motion in this paper. 

The field theoretic method is often employed to study thermodynamic properties of condensed matters. The standard diagramatic technique is 
of course applicable
to treat anharmonicity as the perturbation\cite{cowley}. 
However, in the extremely anharmonic case like ${\rm KOs_{2}O_{6}}$ or tunneling state in which the harmonic part of
the local potential is almost zero or negative, it is not reasonable to start from harmonic potential as the unperturbed state.  
To treat the full effect of anharmonicity, it is necessary for us to develop a new technique because the Wick's theorem is inapplicable anymore.
In this paper, we use the set of eigenstates of effective local ion potential as the starting point, and treat electron-ion interaction as 
the perturbation. In contrast to the harmonic case, the transition energy to the neighboring state is not constant, and
transition can take place even to not neighboring states in an anharmonic potential\cite{hui}. 
Introducing self-energy, vertex-correction, and normalization factor for each transition with various transition energy,
we will show that it is possible to obtain a general expression of 
the Green's function of fully anharmonic lattice vibration within the accuracy of the second order of the electron-ion interaction. Employing that expression,
we will discuss the characteristic properties of spectral functions of the anharmonic oscillator and its application to the NMR relaxation rate, 
and finally comment on its applicability to the study of mass-enhancement and superconductivity.

\section{Model and Calculation of Green's Function}
\subsection{Model}
We consider the system of a local ion oscillator and the conducting paramagnetic electrons 
as described by the following Hamiltonian. For simplicity, we consider a one-dimensional
oscillator.

\begin{equation}
H_0 = \sum_{{\bf k} \sigma} \xi_{{\bf k}} c_{{\bf k}\sigma}^{\dagger}c_{{\bf k}\sigma}  + \sum_{m=0}^{\infty} E_m |m\rangle \langle m| 
\end{equation}

As usual, $c_{{\bf k}\sigma}$ and $ c_{{\bf k}\sigma}^{\dagger}$ are annihilation and creation operators of Fermions, and
$\xi_{{\bf k}}$ denotes a dispersion measured from the chemical potential. $E_m$ and $|m\rangle$ are the set of eigenenergies and orthonormal eigenstates 
of the oscillator in a center-symmetric local potential $V(x)$.
\begin{equation}
\left(-\frac{1}{2M} \frac{d^2}{dx^2} + V(x) \right)|m\rangle  = E_m |m\rangle 
\end{equation}
\begin{equation}
V(x)=\sum_{n=1}^{n_{{\rm max}}} \frac{k_{2n}}{2n}x^{2n}
\end{equation}
To form the bound states, $k_{n_{{\rm max}}}$ must be positive. 

Let us assume that ion displacement couples linearly with electrons, then we write electron-ion interaction as follows.
\begin{equation}
H_{{\rm int}} = \frac{1}{N}\sum_{{\bf kq}\sigma} V({\bf q})  c_{{\bf k}+{\bf q}\sigma}^{\dagger}c_{{\bf k}\sigma} x
\end{equation}
where $N$ is the total number of lattice site, and $V({\bf q})$ is determined from the inner product of the ion displacement direction
and the Fourier transformation of the first derivative of the ion  potential for  the electrons. 

\subsection{Formulation of the Green's function}
We define the general retarded Green's function of lattice vibration as follows.

\begin{equation}
D^R(\omega)=-{\rm i}\int^{\infty}_0 \!\!\!\! {\rm d}t \hspace{1mm} {\rm e}^{{\rm i}\omega t} \langle[x(t),x(0)]\rangle
\end{equation}
Here, $x(t)$ is the Heisenberg representation of the ion displacement, $[A,B]$ denotes the commutation relation, and the bra-ket $\langle \cdots \rangle$ denotes
the statistical average. The Green's function is directly related to the ion polarization.

On the other hand, the temperature Green's function is defined by employing the imaginary time Heisenberg operators.
\begin{eqnarray}
 d(\tau) &=& -\langle T_{\tau}(x(\tau)x(0)) \rangle \nonumber \\
         &=&\left\{
       \begin{array}{l}
            -\langle x(\tau)x(0)\rangle \hspace{5mm}(\beta>\tau>0)\\
            -\langle x(0)x(\tau)\rangle \hspace{5mm}(0>\tau>-\beta)
       \end{array}
           \right. 
\end{eqnarray}

Because of the bosonic definition and the cyclic property of mathematical trace, the temperature Green's function has the 
following periodic property concerning the imaginary time.  
\begin{equation}
d(\tau + \beta) = d(\tau)\hspace{1cm}(\tau<0)
\end{equation}

Then, Fourier expansion is defined using the Matsubara frequency $\omega_k = 2k\pi T$, where $k$ is an integer.
\begin{equation}
d({\rm i}\omega_k) = \int^{\beta}_{0} \!\!\!\! {\rm d}\tau \hspace{1mm} {\rm e}^{{\rm i}\omega_k \tau} d(\tau)
\end{equation}

By changing to the Lehmann representation, we can show that these two functions are related through analytic continuation\cite{AGD}.
\begin{equation}
D^R(\omega) = d(\omega+{\rm i}\delta) \hspace{5mm}(\omega>0,\delta\rightarrow +0)
\end{equation}

This formalism above is completely the same with the usual harmonic phonon.
\subsection{2nd order perturbation}
Now, we calculate $d(\tau)$ for $\tau>0$ considering $H_{{\rm int}}$ as the perturbation.
\begin{equation}
d(\tau>0)= -\frac{{\rm Tr}({\rm e}^{-\beta(H_0 + H_{{\rm int}})}{\rm e}^{(H_0 + H_{{\rm int}})\tau}x{\rm e}^{-(H_0 + H_{{\rm int}})\tau}x)}{{\rm Tr}({\rm e}^{-\beta(H_0 + H_{{\rm int}})})}
\end{equation}

Because the diagonal matrix elements of $x$ are zero for the eigenstates in center symmetric potential, the second order
of $H_{{\rm int}}$ is the lowest contribution. Introducing the interaction representation,
\begin{eqnarray}
&&{\rm e}^{-(H_0 + H_{{\rm int}})\tau} = {\rm e}^{-H_0\tau} S(\tau)\\
&& S(\tau) = 1-\int^{\tau}_{0}{\rm d}\tau_1 \hspace{1mm} {\rm e}^{H_0 \tau_1}H_{{\rm int}}{\rm e}^{-H_0\tau_1} \nonumber\\
&& + \int^{\tau}_{0} \!\!\! \int^{\tau_{1}}_{0}\!\!\!\! {\rm d}\tau_{1} {\rm d}\tau_{2} \hspace{1mm} {\rm e}^{H_0 \tau_1}H_{{\rm int}}{\rm e}^{-(H_0\tau_1-H_0\tau_2)}H_{{\rm int}}{\rm e}^{-H_0\tau_2} \nonumber\\
&&\hspace{1.5cm}+O(H_{{\rm int}}^3)
\end{eqnarray}
our task now is to evaluate the second order contribution.

If the ion potential is harmonic, we can obtain arbitrary order corrections employing the well-known Feynman diagrams.
However, once the potential include anharmonicity, we can not employ this technique anymore. The Feynman diagram
is based on the Wick's theorem, which shows that the statistical average of the product of interaction 
representation operators can be decomposed into the product of the 
statistical average of various combinations of two operators\cite{FW}. To prove this theorem,
we need the following property of the interaction representation operator, that is we can factor out the 
time dependent exponential as the classical number like ${\rm e}^{\omega_0 \tau a^{\dagger}a}a{\rm e}^{-\omega_0 \tau a^{\dagger}a}= {\rm e}^{-\omega_0 \tau}a$. 
However, we can not use this relation if $H_{0}$ includes anharmonic terms.
Considering this fact, we now collect second order contribution directly without using Feynman diagram technique. 

After some algebra, we obtain a general expression for the second order corrections.
\begin{eqnarray}
&&\hspace{-1cm}\delta^{(2)} d(\tau)=\frac{2}{N^2}\sum_{{\bf kq}}|V({\bf q})|^2\int^{\beta}_{\tau}\!\!\! \int^{\tau_1}_{\tau}\!\!\!\! 
{\rm d}\tau_1{\rm d}\tau_2\hspace{1mm}{\rm F}(\tau_1,\tau_2,\tau)\nonumber\\
              &&\times g^{(0)}_{{\bf k}}(\tau_1-\tau_2)g^{(0)}_{{\bf k}+{\bf q}}(\tau_2-\tau_1)\label{exact1}\\
&&+\frac{2}{N^2}\sum_{{\bf kq}}|V({\bf q})|^2\int^{\tau}_{0}\!\!\! \int^{\tau_1}_{0}\!\!\!\! {\rm d}\tau_1{\rm d}\tau_2\hspace{1mm}{\rm F}(\tau,\tau_1,\tau_2)\nonumber\\
              &&\times g^{(0)}_{{\bf k}}(\tau_1-\tau_2)g^{(0)}_{{\bf k}+{\bf q}}(\tau_2-\tau_1)\label{exact2}\\
&&+\frac{2}{N^2}\sum_{{\bf kq}}|V({\bf q})|^2\int^{\beta}_{\tau}\!\!\! \int^{\tau}_{0}\!\!\!\! {\rm d}\tau_1{\rm d}\tau_2\hspace{1mm}{\rm F}(\tau_1,\tau,\tau_2)\nonumber\\
              &&\times g^{(0)}_{{\bf k}}(\tau_1-\tau_2)g^{(0)}_{{\bf k}+{\bf q}}(\tau_2-\tau_1)\label{exact3}\\
&&-\frac{2}{N^2}\sum_{{\bf kq}}|V({\bf q})|^2\int^{\beta}_{0}\!\!\! \int^{\tau_1}_{0}\!\!\!\! {\rm d}\tau_1{\rm d}\tau_2\hspace{1mm} 
d^{(0)}(\tau)d^{(0)}(\tau_1-\tau_2)\nonumber\\
              &&\times g^{(0)}_{{\bf k}}(\tau_1-\tau_2)g^{(0)}_{{\bf k}+{\bf q}}(\tau_2-\tau_1)\label{exact4}
\end{eqnarray}

The free ion and electron propagators are  written explicitly as follows for positive $\tau$,
where $Z_{i0}$ denotes the partition function of free ion part, and $f(\xi)$ denotes the Fermi distribution function $1/({\rm e}^{\xi/T}+1)$. 
\begin{eqnarray}
d^{(0)}(\tau)&\hspace{-3mm}=&\hspace{-4mm}-\langle x(\tau)x(0)\rangle_0 \nonumber \\
           &\hspace{-3mm}=&\hspace{-4mm}-Z_{i0}^{-1}\sum_{mn}{\rm e}^{-\beta E_{m}}{\rm e}^{(E_{m}-E_{n})\tau}|\langle m|x|n \rangle|^2\\
&&\hspace{-23mm}g^{(0)}_{{\bf k}}(\tau)=-\langle c_{\bf k}(\tau)c_{\bf k}^{\dagger}(0)\rangle_0=-{\rm e}^{-\xi_{{\bf k}} \tau}\left(1-f(\xi_{{\bf k}})\right)\\
&&\hspace{-23mm}g^{(0)}_{{\bf k}+{\bf q}}(-\tau)=+\langle c_{{\bf k}+{\bf q}}^{\dagger}(0)c_{{\bf k}+{\bf q}}(\tau)\rangle_0={\rm e}^{\xi_{{\bf k}+{\bf q}}\tau}f(\xi_{{\bf k}+{\bf q}})
\end{eqnarray}

The function ${\rm F}(\lambda,\mu,\nu)$ is defined as follows. 
\begin{eqnarray}
 &&\hspace{-5mm}{\rm F}(\lambda,\mu,\nu)=Z_{i0}^{-1}\sum_{i_1 \sim i_4}
\langle i_1|x|i_2 \rangle\langle i_2|x|i_3 \rangle\langle i_3|x|i_4 \rangle\langle i_4|x|i_1 \rangle \nonumber \\ 
 &&\times {\rm e}^{\left(-\beta E_{i_1} + (E_{i_1}-E_{i_2})\lambda + (E_{i_2}-E_{i_3})\mu +(E_{i_3}-E_{i_4})\nu \right)} \label{F}
\end{eqnarray}

Now, we introduce diagrams which is different from the Feynman diagrams but
give some hints to understand the physical meaning of each contribution, Eq(\ref{exact1})$\sim$(\ref{exact4}).
The rules of drawing the diagrams and correspondence with equations are the following.
\begin{enumerate}
\item There are two kinds of vertices, one of which is the external vertex with external imaginary time $0$ or $\tau$, and the other is the 
internal vertex with internal imaginary time $\tau_1$ or $\tau_2$.
We arrange them so that attached imaginary time arguments proceed clockwise, and connect them by wavy lines
with suffixes of ion eigenstates. Especially for the vertex with the minimum imaginary 
time argument, we also attach the inverse temperature. 
\item We connect internal vertices with two oppositely directed arrows which denote the electron-hole excitations.
\item We assign x-matrix elements between two ion eigenstates sandwiching each vertex.
\item Corresponding to each vertex, we multiply by the exponential factor in which the imaginary time is multiplied by 
the eigenenergy in smaller imaginary time region with negative sign, and the other one is multiplied with positive sign. 
\item For two arrows, we assign free electron lines with certain wave vectors with imaginary time arguments with opposite
signs.
\item Defining the integer $l$, which is the number of the connected diagrams, we multiply $(-1)^{l+1}Z_{i0}^{-l}$. 
\item We multiply $|V({\bf q})|^2/N^2$ which denotes that the interaction accompanies $\pm {\bf q}$ momentum transfer at two internal vertices, 
and the factor 2 denoting the spin summation,
and finally take summation over two wave vectors and ion eigenstates.
\end{enumerate}

\begin{figure}[h]
\begin{center}
\begin{tabular}{cc}
 \begin{minipage}{0.5\hsize}
\subfigure[]{\includegraphics[width=\hsize]{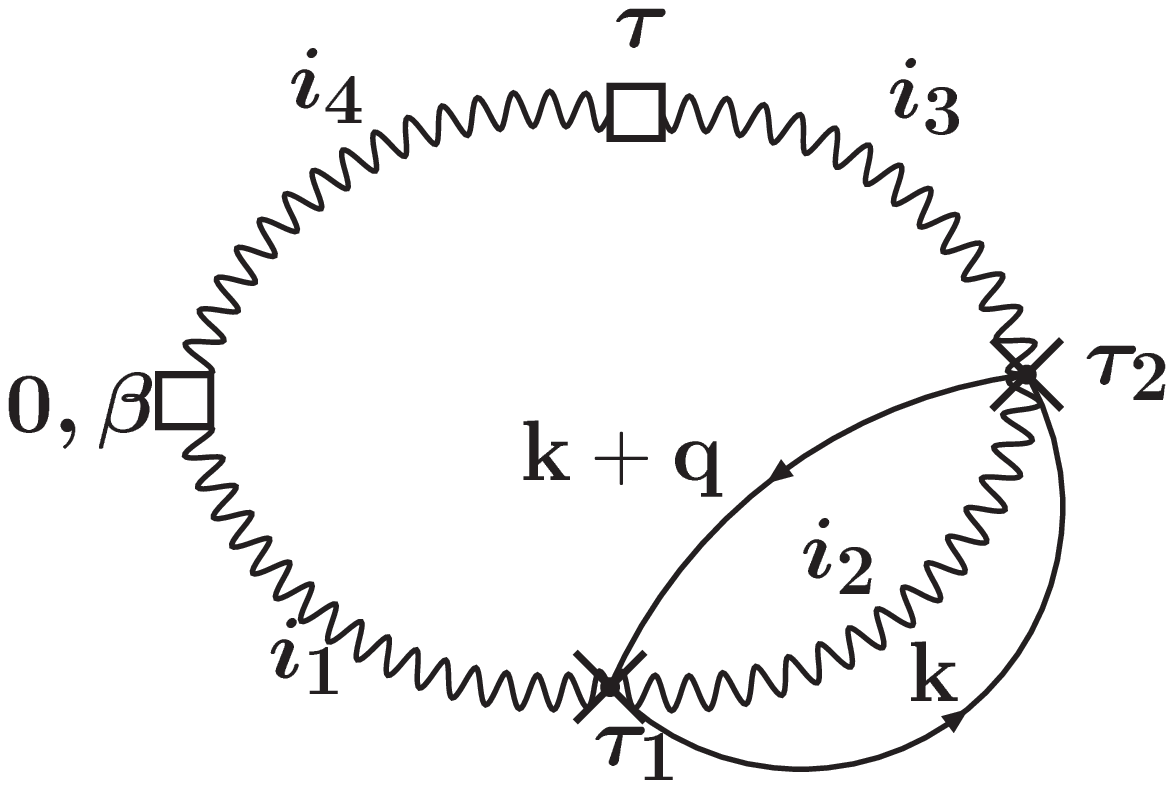}
   \label{diagram1}}
 \end{minipage} 
\hfill
 \begin{minipage}{0.5\hsize}
\subfigure[]{\includegraphics[width=\hsize]{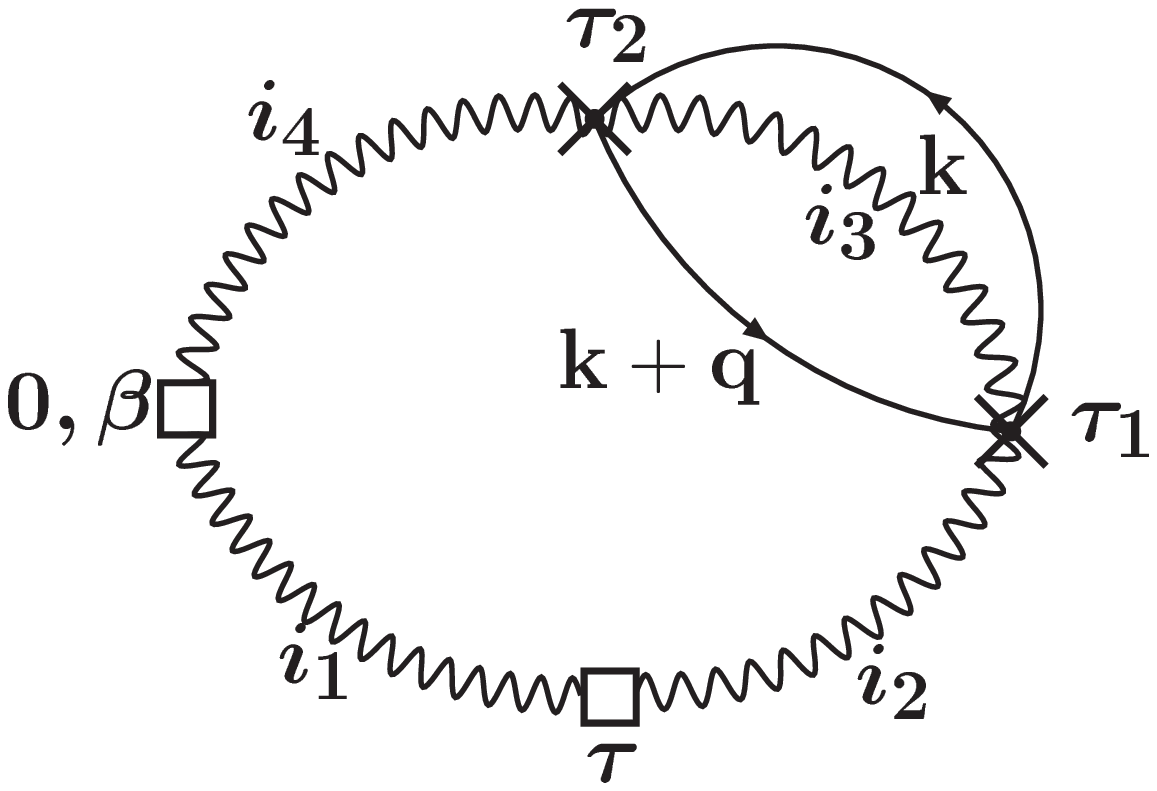}
   \label{diagram2}}
 \end{minipage}\\
 \begin{minipage}{0.5\hsize}
\subfigure[]{\includegraphics[width=\hsize]{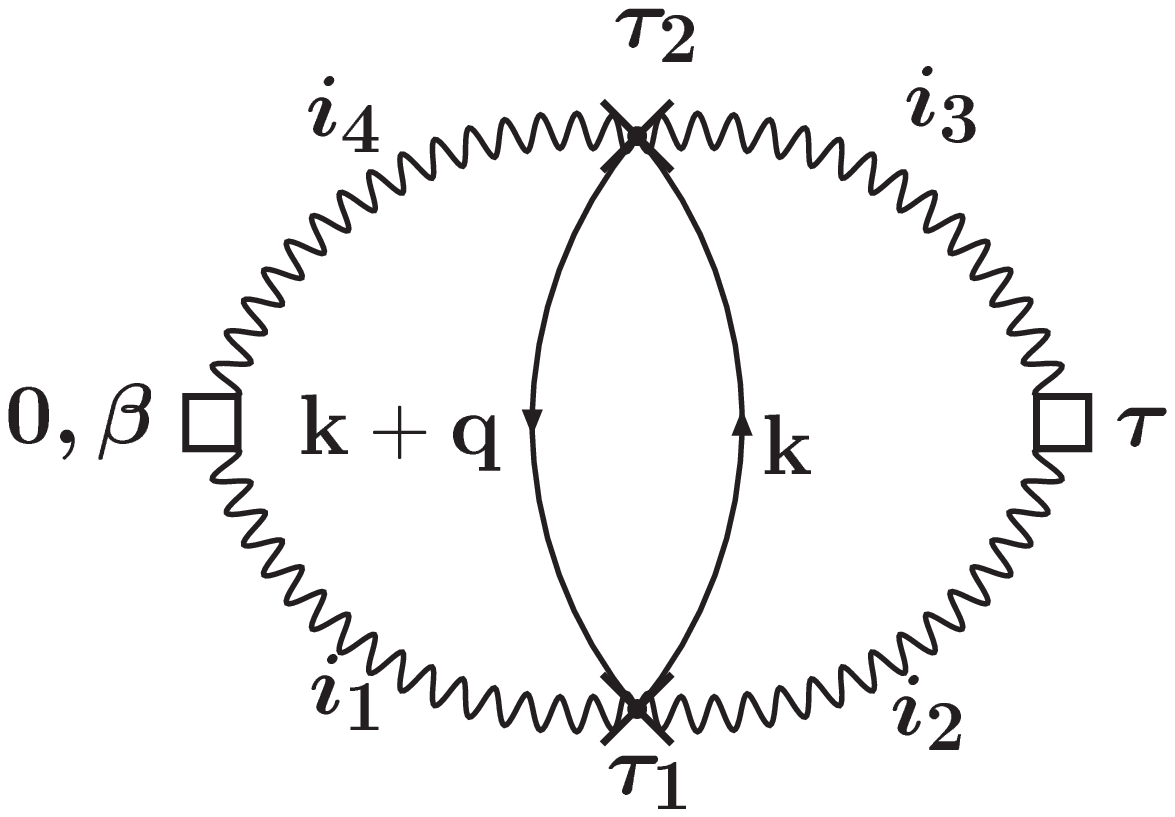}
   \label{diagram3}}
 \end{minipage} 
\hfill
 \begin{minipage}{0.5\hsize}
 \subfigure[]{\includegraphics[width=\hsize]{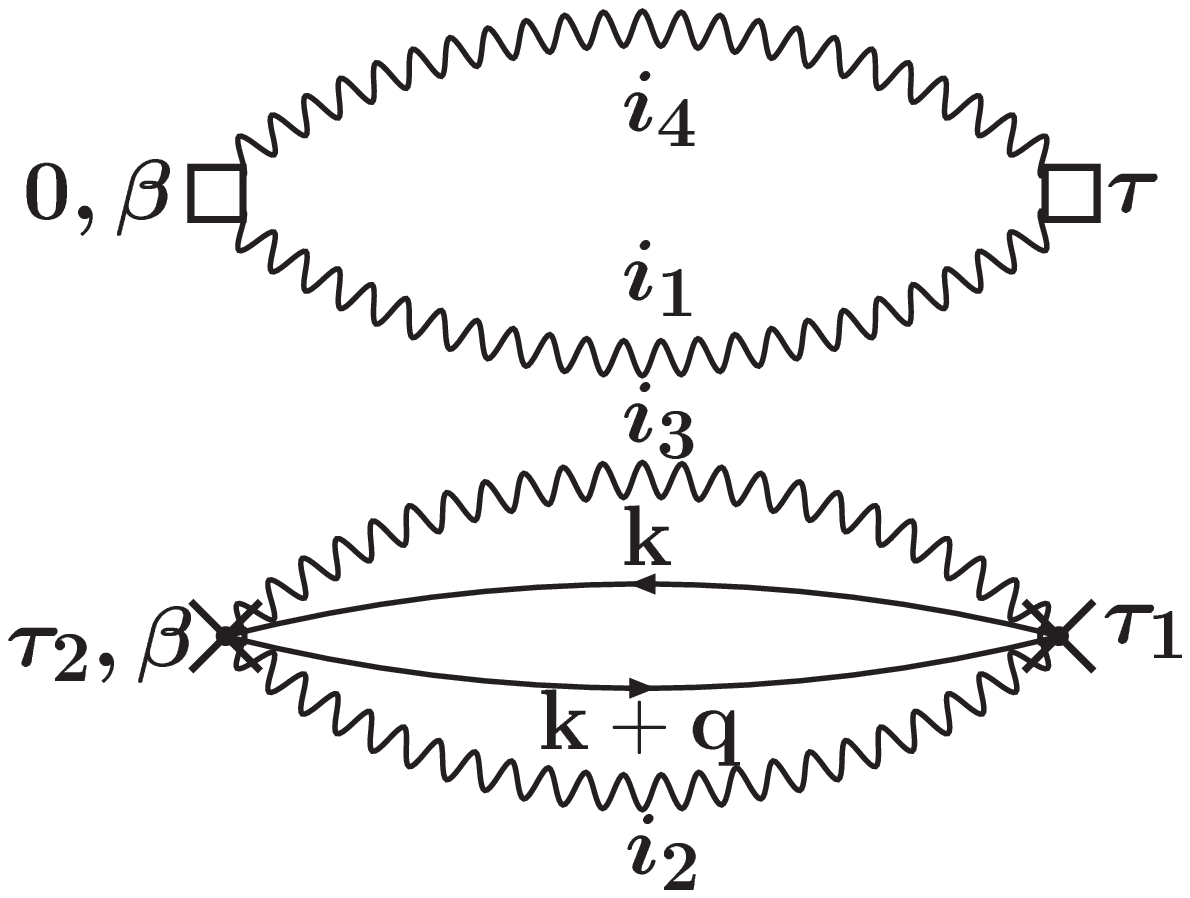}
   \label{disc}}
 \end{minipage}
\end{tabular} 
\end{center}
\caption{Diagramatic representations of the four contributions, Eq.(\ref{exact1})$\sim$(\ref{exact4}).
The details of the notations are written in the main text.}
\label{diagrams}
\end{figure}

Following this rule, we can draw diagrams as shown in Figs.\ref{diagram1}-\ref{disc} which correspond to Eq.(\ref{exact1})$\sim$(\ref{exact4}), respectively.
\subsection{\label{sec:main}Self-energy, vertex-correction, normalization factor}
\begin{figure}[h]
\begin{center}
\begin{tabular}{cc}
 \begin{minipage}{0.4\hsize}
\subfigure[]{\includegraphics[width=\hsize]{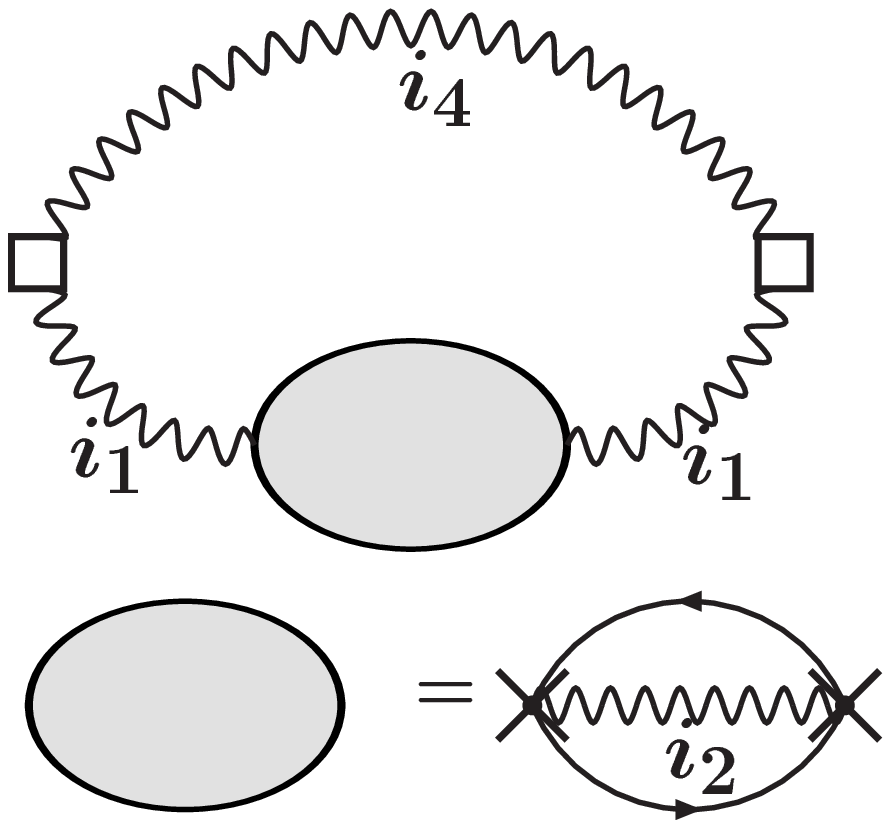}
   \label{se1}}
 \end{minipage} 
\hfill
 \begin{minipage}{0.4\hsize}
\subfigure[]{\includegraphics[width=\hsize]{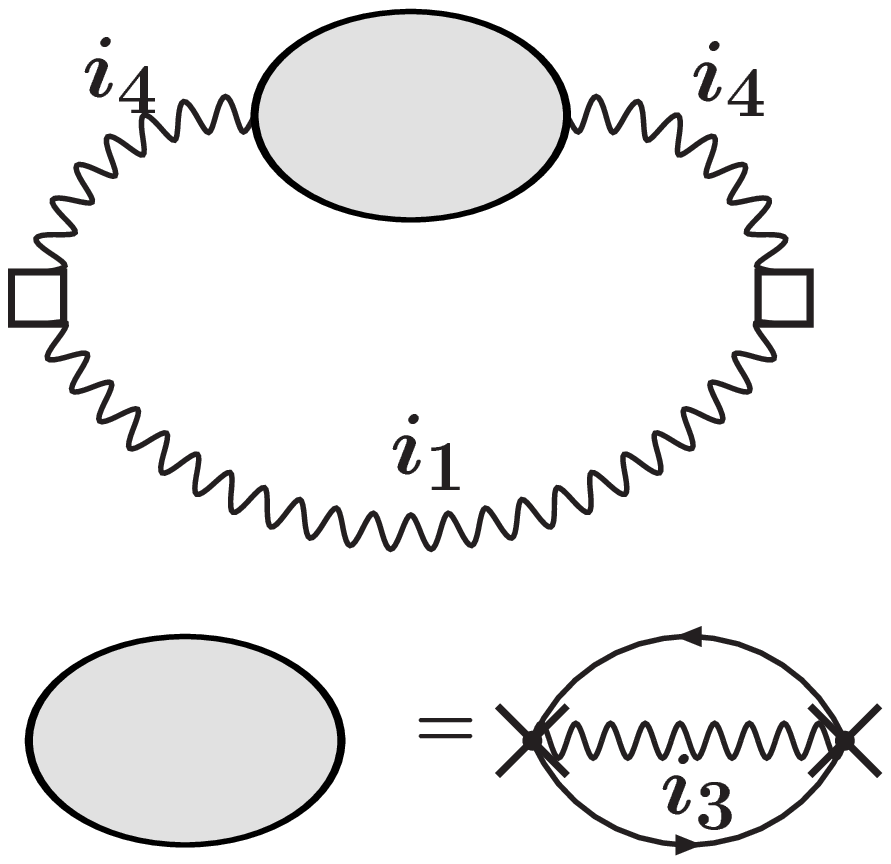}
   \label{se2}}
 \end{minipage}
\end{tabular}
\end{center}
\caption{Two kinds of self-energy type contributions originating from Fig.\ref{diagram1} and Fig\ref{diagram2} respectively.}
\end{figure}

Now we use the Fourier expansion, and arrange various contributions by taking care of the physical
meanings indicated by drawing diagrams.
First of all, we consider Eq.(\ref{exact1}) which corresponds to Fig.\ref{diagram1}.
If $i_1=i_3$ in Fig.\ref{diagram1}, we can rewrite these contributions into the form of Fig.\ref{se1}. The gray circle
may be treated as the self-energy for the $i_1-i_4$ transition. Its real part gives renormalization of the transition
energy, and the imaginary part gives its life-time. The same thing is true in Fig.\ref{diagram2} with  $i_2=i_4$
as shown in Fig.\ref{se2}. In the following, we call this type of contributions ``self-energy type''.
After some calculations, we can see that all the self-energy type contributions can be divided into two groups,
one of which has $({\rm i}\omega_n+E_{i_1}-E_{i_4})^{-1}$ as a common factor while the other $({\rm i}\omega_n+E_{i_1}-E_{i_4})^{-2}$.

\begin{figure}[h]
\begin{center}
\begin{tabular}{ccc}
 \begin{minipage}{0.33\hsize}
\subfigure[]{\includegraphics[width=\hsize]{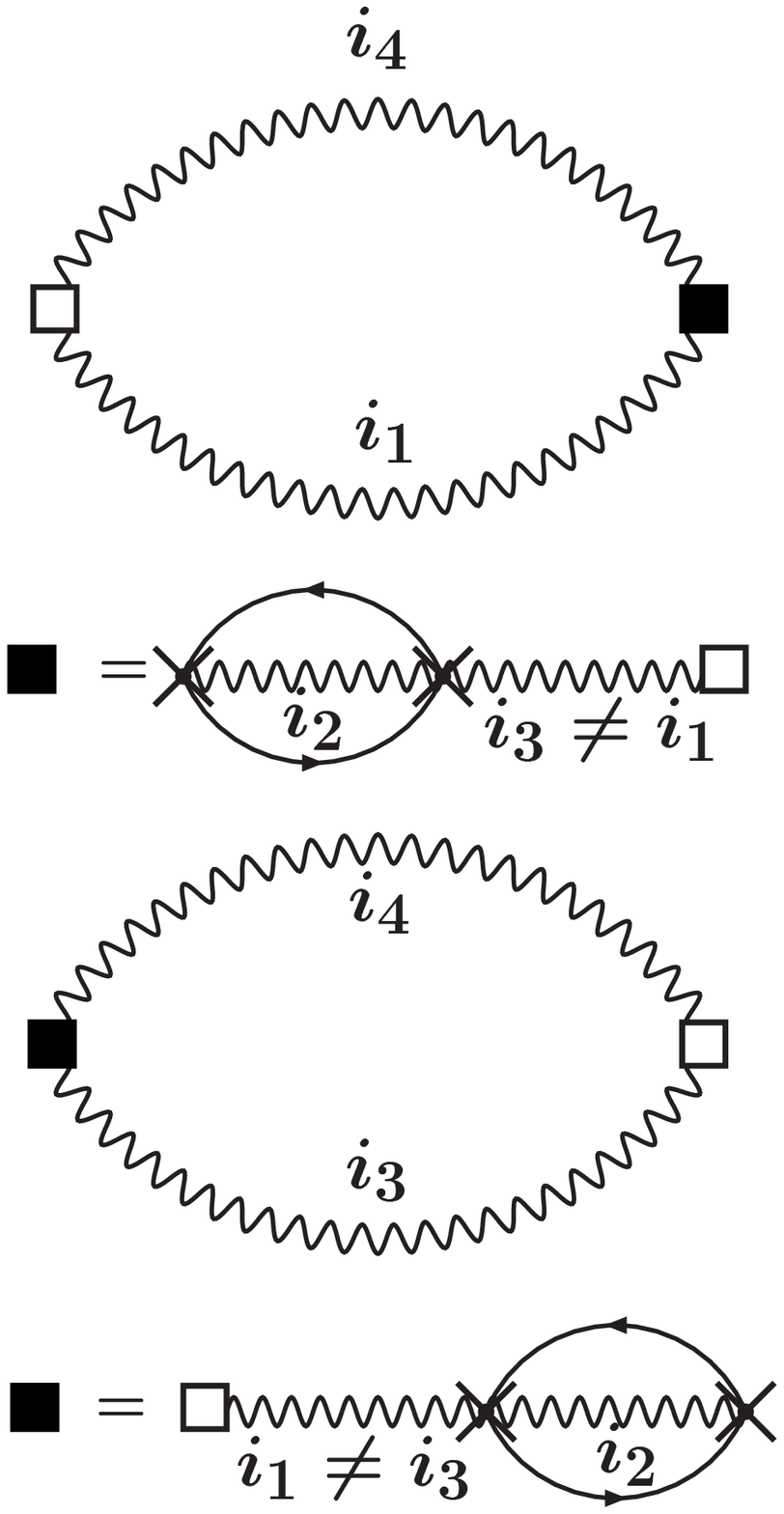}
   \label{vt1}}
 \end{minipage} 
\hfill
 \begin{minipage}{0.33\hsize}
\subfigure[]{\includegraphics[width=\hsize]{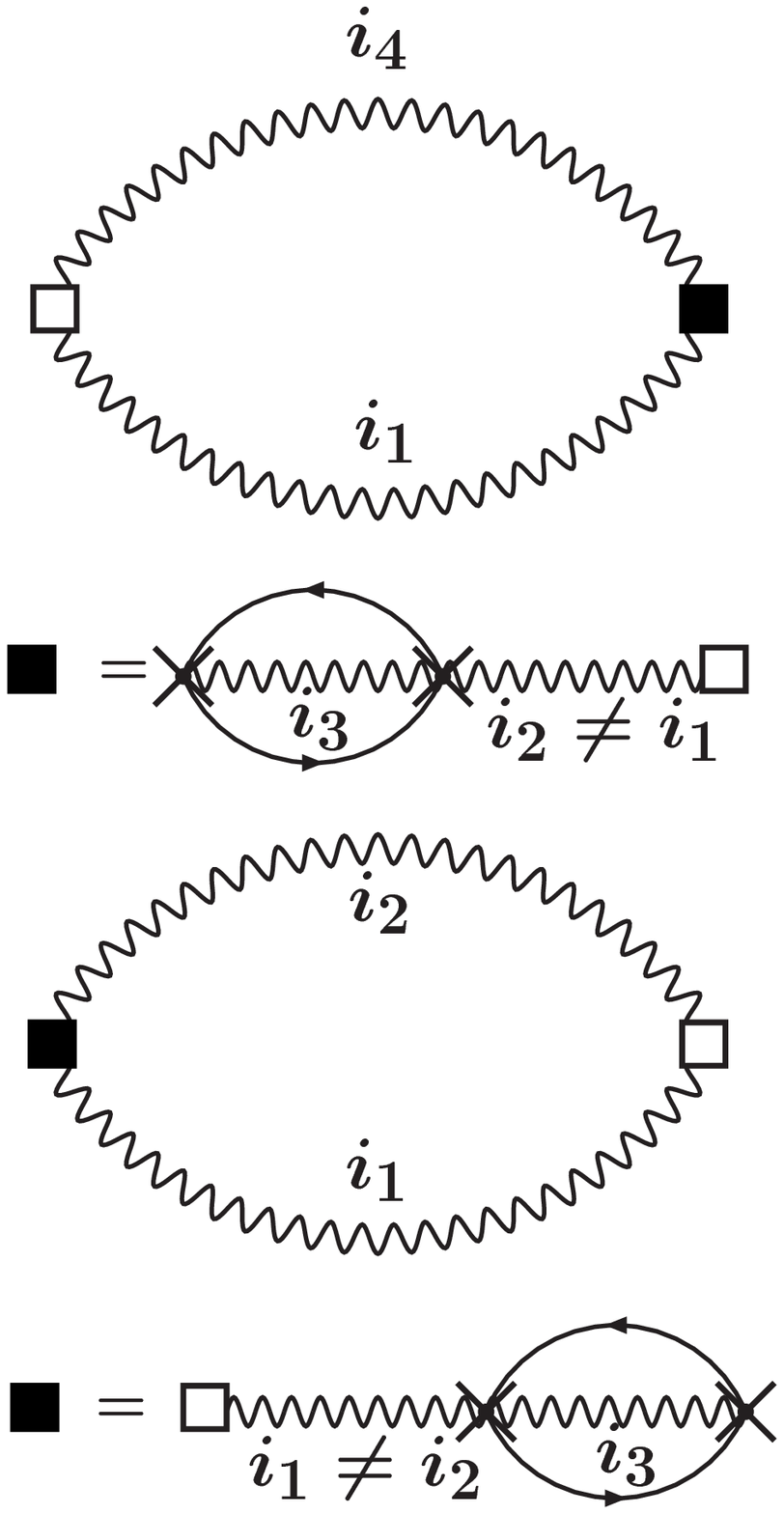}
   \label{vt2}}
 \end{minipage}
\hfill
 \begin{minipage}{0.33\hsize}
\subfigure[]{\includegraphics[width=\hsize]{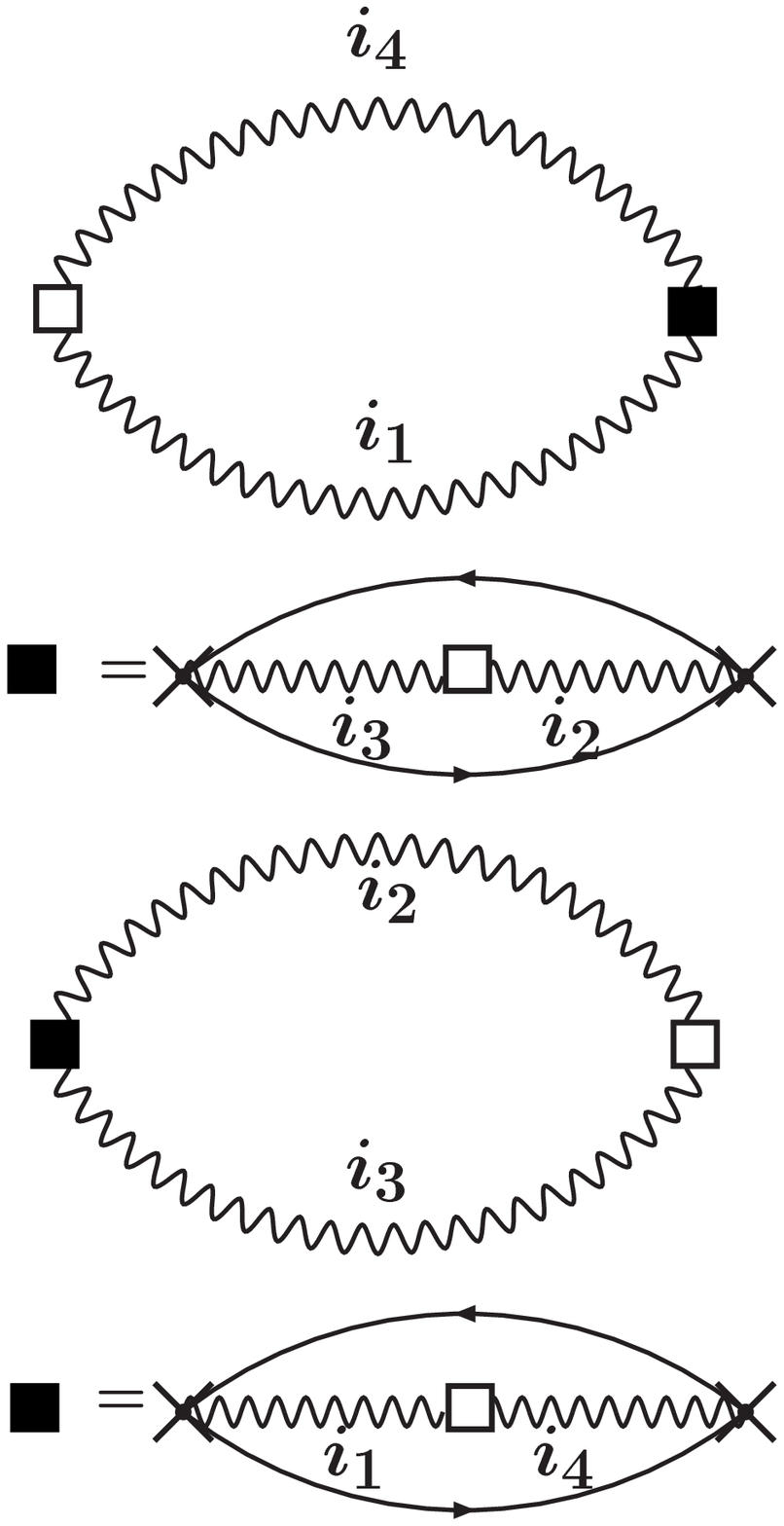}
   \label{vt3}}
 \end{minipage}
\end{tabular} 
\end{center}
\caption{Three kinds of vertex-correction type contributions. See the main text for more details.}
\end{figure}

Next we consider the following three cases, Fig.\ref{diagram1} with $i_1\neq i_3$, Fig.\ref{diagram2}
with $i_2\neq i_4$, and any type of processes in Fig.\ref{diagram3}. We can regard these contributions
as a modification of matrix elements of $x$ and call ``vertex-correction type''.
For example concerning Fig.\ref{diagram1}, one can regard this contribution in two ways as shown in Fig.\ref{vt1}.
The external vertex with imaginary time $\tau$ and two internal vertices combined by two
wavy lines are united and considered as one renormalized vertex which is specified by the thick square in the top figure while it is also possible to combine the other
external vertex with two internal vertices into one renormalized vertex as in the bottom figure. One can regard
the top one as the correction to the $i_1$-$i_4$ transition while the other to $i_3$-$i_4$ transition. Because the product
of matrix elements of $x$ in the function ${\rm F}$ has $i_1$-$i_3$ permutation symmetry, it is always possible to re-interpret
$i_3$-$i_4$ type contributions as $i_1$-$i_4$ type one by permuting $i_1$ and $i_3$. In the course of calculation, we decompose every term
of Fourier expansion coming from Eq.(\ref{exact1}) into partial fractions and divide into two groups, one of which
has $({\rm i}\omega_n+E_{i_1}-E_{i_4})^{-1}$ as a common factor and the other $({\rm i}\omega_n+E_{i_3}-E_{i_4})^{-1}$. About the latter one,
we permute $i_3$ and $i_1$ so that all the terms have the common factor $({\rm i}\omega_n+E_{i_1}-E_{i_4})^{-1}$ which correspond to
the correction to the $i_1$-$i_4$ transition. Similarly, Fig.\ref{vt2} and \ref{vt3} indicate two ways of consideration
of Fig.\ref{diagram2} with $i_2\neq i_4$ and Fig.\ref{diagram3} respectively. In both cases, we follow the similar procedure
mentioned above. When we decompose the Fourier expansion of Eq.(\ref{exact2}) and Eq.(\ref{exact3}) into partial fractions,
we notice that all terms can be divided into two groups in each case. From Eq.(\ref{exact2}), we obtain two groups with two kinds of
common factor $({\rm i}\omega_n+E_{i_1}-E_{i_4})^{-1}$ or $({\rm i}\omega_n+E_{i_1}-E_{i_2})^{-1}$. About the latter one, we permute $i_2$ and $i_4$ considering the
permutation symmetry of the product of matrix elements of $x$ in the function ${\rm F}$. On the other hand from Eq.(\ref{exact3}), we
will obtain two groups with $({\rm i}\omega_n+E_{i_1}-E_{i_4})^{-1}$ or $({\rm i}\omega_n+E_{i_2}-E_{i_3})^{-1}$
as a common factor. We permute $i_2$ and $i_1$, $i_3$ and $i_4$ simultaneously
concerning the latter one.
Eventually, we can arrange ``vertex-correction type'' so that all the terms have the common factor $({\rm i}\omega_n+E_{i_1}-E_{i_4})^{-1}$ which 
correspond to the correction to $i_1$-$i_4$ transition.  
 
The remaining part of Eq.(\ref{exact4}) which is represented by the ``disconnected diagram'' Fig.\ref{disc} will contribute just as the total normalization of $d^{(0)}$. 

Now, we define the self-energy $\Pi$, vertex correction $\delta x$, and normalization factor $\Lambda$ for each transition.
\begin{eqnarray}
&&\hspace{-5mm}d({\rm i}\omega_n)
=Z_{i0}^{-1}\sum_{i_1i_4}({\rm e}^{-\beta E_{i_1}}-{\rm e}^{-\beta E_{i_4}}) \nonumber \\
&&\hspace{-2mm}\times \frac{|\langle i_1|x|i_4 \rangle|^2(1+\Lambda(i_1,i_4))
+\delta x({\rm i}\omega_n,i_1,i_4)}{{\rm i}\omega_n+E_{i_1}-E_{i_4}-\Pi({\rm i}\omega_n,i_1,i_4)} \label{svn}
\end{eqnarray}

Within the accuracy of the second order, this form is equal to the following Taylor expansion.
\begin{eqnarray}
&&\hspace{-13mm}d({\rm i}\omega_n)=d^{(0)}({\rm i}\omega_n)\nonumber \\
&&\hspace{-12.5mm}+Z_{i0}^{-1}\sum_{i_1i_4}({\rm e}^{-\beta E_{i_1}}-{\rm e}^{-\beta E_{i_4}})
\frac{|\langle i_1|x|i_4 \rangle|^2\Pi({\rm i}\omega_n,i_1,i_4)}{({\rm i}\omega_n+E_{i_1}-E_{i_4})^{2}} \label{selfenergy}\\
&&\hspace{-12.5mm}+Z_{i0}^{-1}\sum_{i_1i_4}({\rm e}^{-\beta E_{i_1}}-{\rm e}^{-\beta E_{i_4}})\frac{\delta x({\rm i}\omega_n,i_1,i_4)}
{{\rm i}\omega_n+E_{i_1}-E_{i_4}}  \label{vertexcorrection}\\
&&\hspace{-12.5mm}+Z_{i0}^{-1}\sum_{i_1i_4}({\rm e}^{-\beta E_{i_1}}-{\rm e}^{-\beta E_{i_4}})\frac{|\langle i_1|x|i_4 \rangle|^2\Lambda(i_1,i_4)} 
{{\rm i}\omega_n+E_{i_1}-E_{i_4}} \label{normalization}
\end{eqnarray}

We can derive the expressions of $\Pi$, $\delta x$, and $\Lambda$ by the following procedure.
About the self-energy $\Pi({\rm i}\omega_n,i_1,i_4)$, we collect the terms with the common factor $({\rm i}\omega_n+E_{i_1}-E_{i_4})^{-2}$ in the self-energy type
corrections and take them equal to Eq.(\ref{selfenergy}). Next, we collect all the terms of vertex-correction type
as the corresponding part of Eq.(\ref{vertexcorrection}), then
we can obtain the expression of $\delta x({\rm i}\omega_n,i_1,i_4)$.
Finally about the normalization factor $\Lambda(i_1,i_4)$, we collect disconnected diagram contribution and $({\rm i}\omega_n+E_{i_1}-E_{i_4})^{-1}$
liner terms of self-energy type corrections and take them equal to Eq.(\ref{normalization}). The final results are shown in the
Appendix explicitly.

There are some notes worth paying attention to concerning the normalization factor.
In the harmonic case, we can show that the disconnected diagram contribution and $\beta$ linear terms of the self-energy type corrections exactly cancel out
when we take summation over ion eigenstates. This cancellation corresponds to the linked cluster theorem in the Feynman diagramatic technique. 
Additionally, this cancellation also occurs in the anharmonic case if we take the low-temperature limit. Considering this fact, it is reasonable to introduce
normalization factor in the numerator in Eq.(\ref{svn}) and include these terms in $\Lambda$ simultaneously. About the $\beta$ nonlinear terms with 
the common factor $({\rm i}\omega_n+E_{i_1}-E_{i_4})^{-1}$ which comes from the self-energy type corrections originally, it may be ambiguous
whether they should be treated as the self-energy or normalization factor although these two ways are equivalent in the second order
perturbation. Actually, those terms are $\omega$ independent and gives only an unimportant corrections unless the perturbation fails.
Then this ambiguity is not a serious problem.

At the end of this section, we would like to make some comments of the preceding study by Dahm and one of the authors\cite{Dahm2007}.
In the paper, they treated the square-well potential as an example of extremely anharmonic case, and
drew spectrums introducing self-energy phenomenologically which is common to all the transition energies. They succeeded to describe the spectral weight shift
arising from the anharmonicity, but the obtained spectral function includes
unphysical zero points which correspond to the singularities
between various transition energies in the free
propagator. By introducing the self-energy, vertex-correction, and normalization factor for ``each transition'' as discussed above,
we can avoid the unphysical phenomena.
\section{Spectral Function and NMR Relaxation Rate}
We draw spectrums for several typical cases using the retarded Green's function which is obtained by
the analytic continuation.
\begin{equation}
A(\omega)=-\frac{1}{\pi |\langle 0|x|1 \rangle|^2}{\rm Im}D_R(\omega)
\end{equation}

Introducing appropriate scale of length $x=d{\tilde x}$ and ion energy scale $E$, we write the local ion Hamiltonian
as follows.
\begin{equation}
H_{{\rm ion}}=E\left(-\frac{d^2}{d {\tilde x}^2}+a{\tilde x}^2+b{\tilde x}^4\right) \label{local}
\end{equation}
We take the constant $E=10$ as an typical energy scale of Einstein frequency with milli-electron volt in mind as the units,
and change two parameters $a$ and $b$. If $a \geq 0$, we keep the relation $a+b=1$ so that
the potential energy $a{\tilde x}^2+b{\tilde x}^4$ in the units takes a value of unity at ${\tilde x}=1$. If $a<0$, we change only $a$ keeping
$b=1$. Especially if $a=1$ and $b=0$,
the Einstein frequency of this harmonic oscillator is $20$.

Later on, we neglect the momentum dependence of $|V({\bf q})|$ and replace by a constant value $g$ for brevity of calculation. We also
assume the constant density of states.
Except for $\beta$ nonlinear terms in the normalization factor, we can perform the summation over wave-vectors analytically at absolute zero temperature.
We neglect temperature dependence of those terms and employ absolute-zero analytic expression
because the energy scale of electrons is much larger than that of the ion. 
We choose parameters, half of the bandwidth(Fermi energy measured from
the band-bottom)$=100$ keeping large mass-enhancement in mind, and coupling constant and scale of 
length $dg=10$. In the following calculation, we will neglect $\beta$ nonlinear terms in
the normalization factor which will give only a few percent correction with these parameters since
it is time-consuming to perform numerically the integration involved. 

\begin{figure}[h]
\begin{center}
  \includegraphics[width=0.7\hsize,angle=-90]{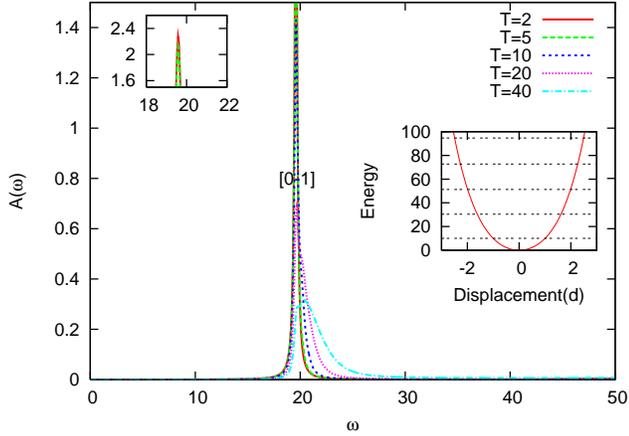}
   \caption{The spectral function of a local ion oscillator(Eq.(\ref{local})) for the parameters (a,b)=(0.98,0.02).
The inset at the top complement the lowest energy excitation peak, and the other one shows the potential and eigenenergies. The eigenstate
suffix runs from 0. For each peak, two numbers are attached to specify transition.}
   \label{0.98_0.02}
\end{center}
\end{figure}
\begin{figure}[h]
\begin{center}
  \includegraphics[width=0.7\hsize,angle=-90]{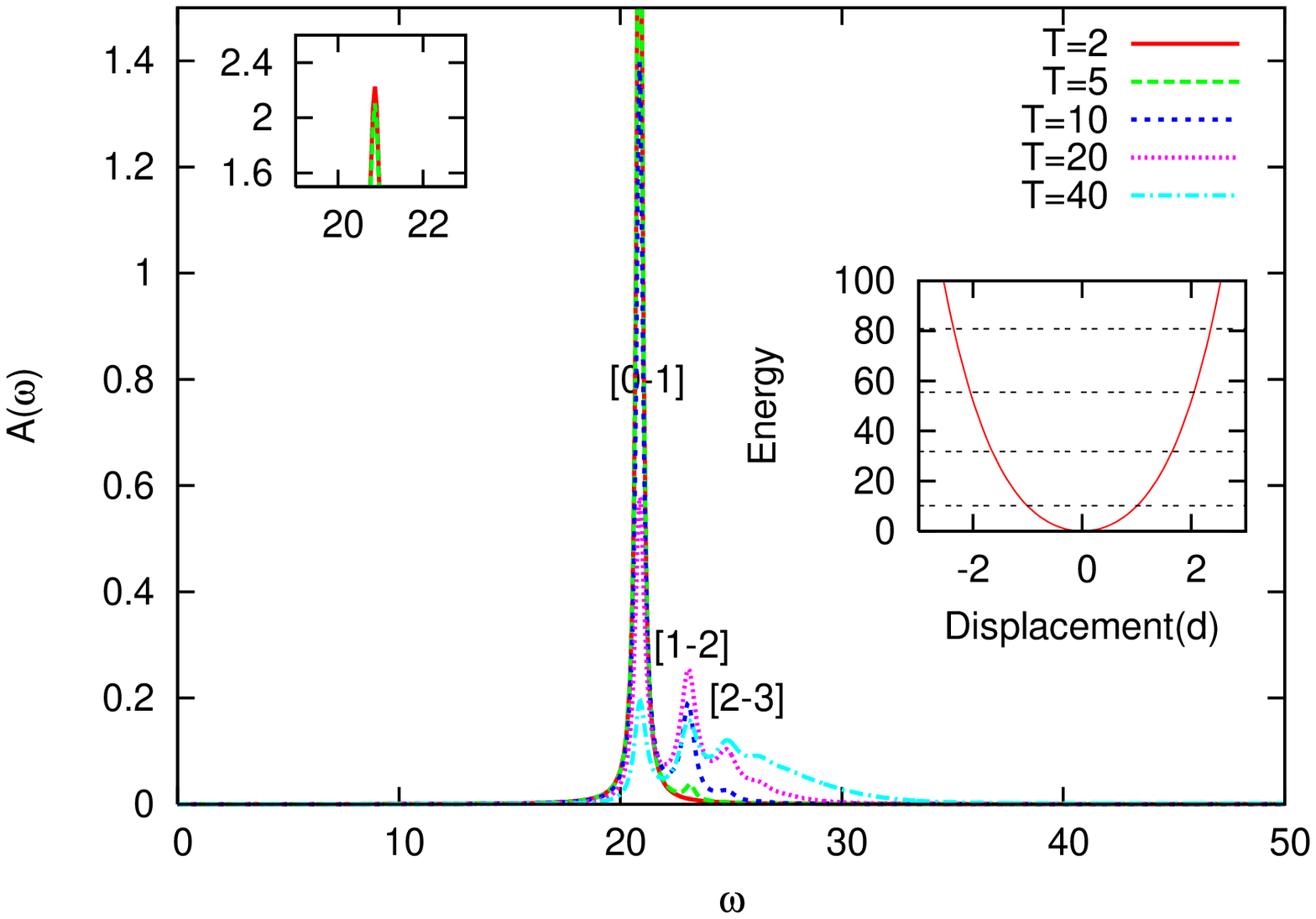}
   \caption{(a,b)=(0.9,0.1)$\hspace{3mm}$The notations used in the figure are the same as Fig.\ref{0.98_0.02}}
   \label{0.9_0.1}
\end{center}
\end{figure}
\begin{figure}[h]
\begin{center}
  \includegraphics[width=0.7\hsize,angle=-90]{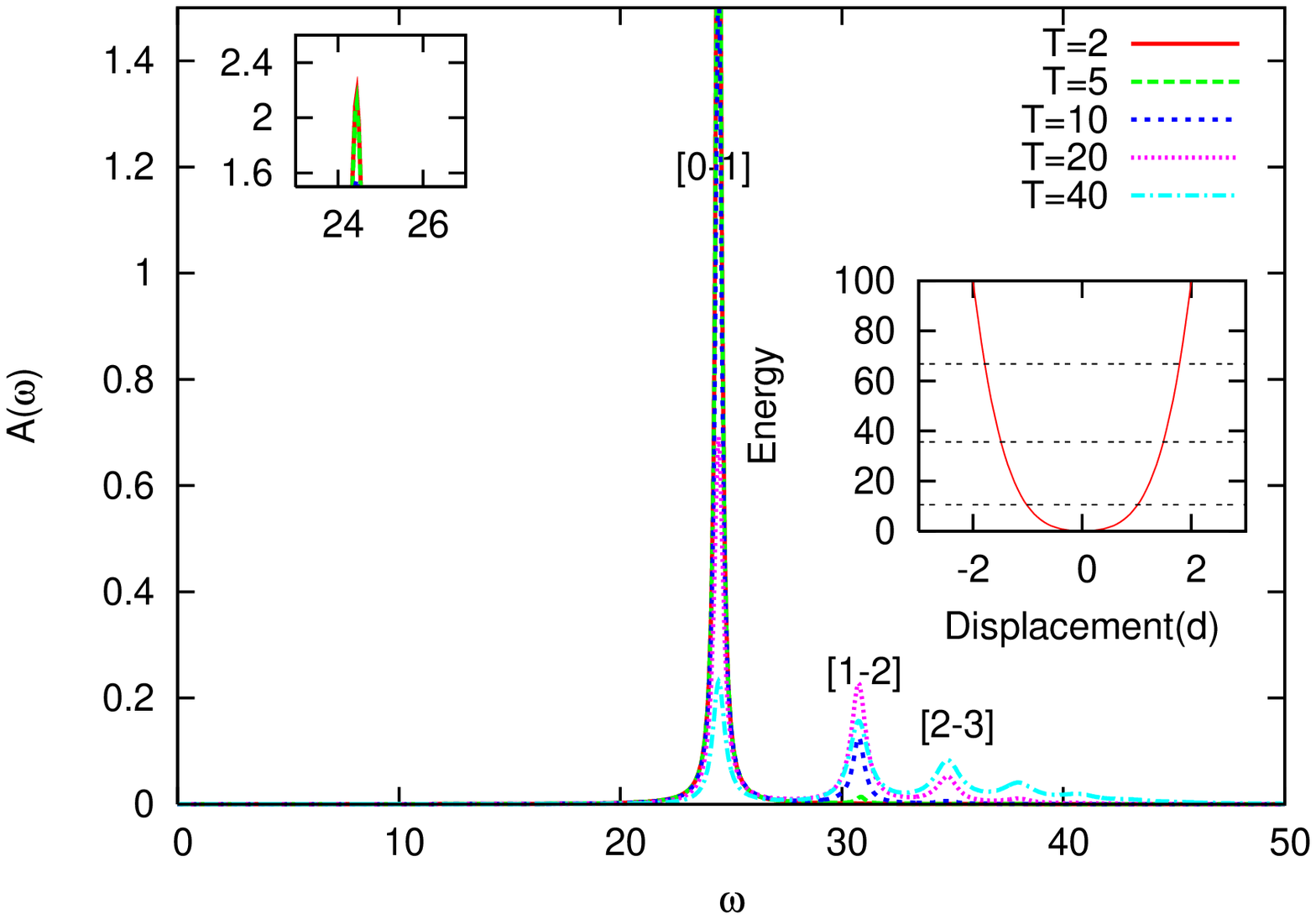}
   \caption{(a,b)=(0.5,0.5)$\hspace{3mm}$The notations used in the figure are the same as Fig.\ref{0.98_0.02}}
   \label{0.5_0.5}
\end{center}
\end{figure}
\begin{figure}[h]
\begin{center}
  \includegraphics[width=0.7\hsize,angle=-90]{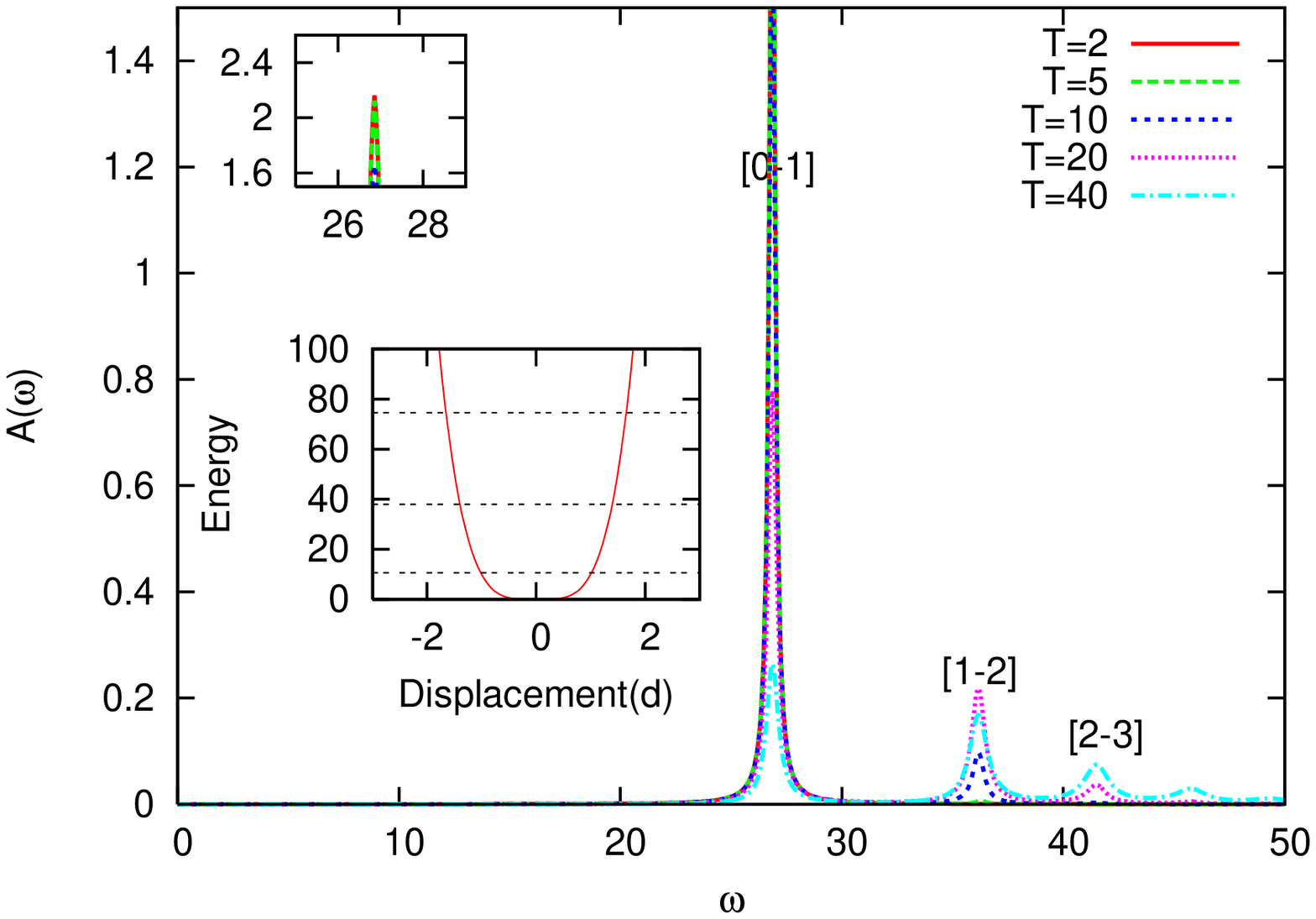}
   \caption{(a,b)=(0,1)$\hspace{3mm}$The notations used in the figure are the same as Fig.\ref{0.98_0.02}}
   \label{0_1}
\end{center}
\end{figure}
\begin{figure}[h]
\begin{center}
  \includegraphics[width=0.7\hsize,angle=-90]{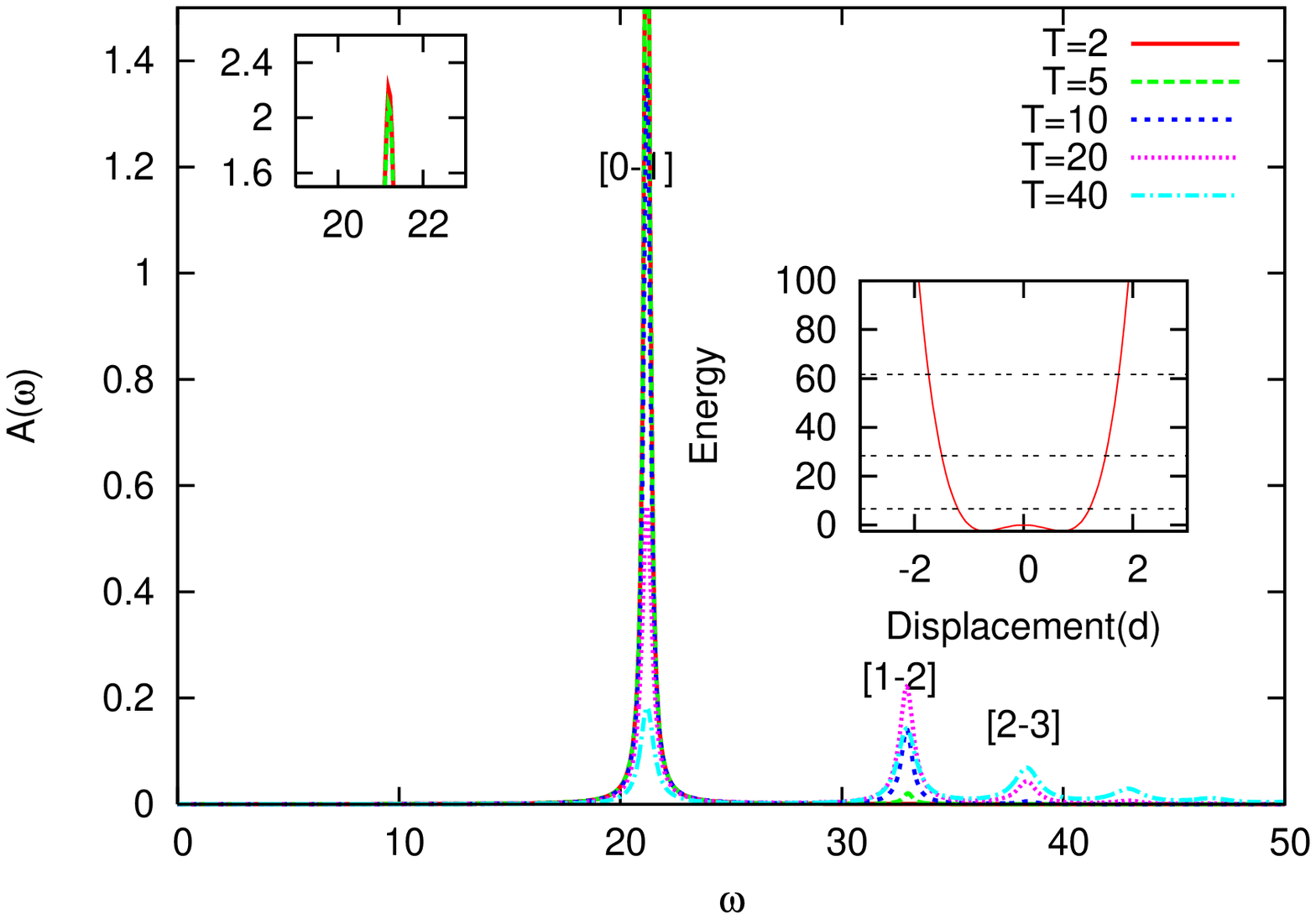}
   \caption{(a,b)=(-1,1)$\hspace{3mm}$The notations used in the figure are the same as Fig.\ref{0.98_0.02}}
   \label{m1_1}
\end{center}
\end{figure}
\begin{figure}[h]
\begin{center}
  \includegraphics[width=0.7\hsize,angle=-90]{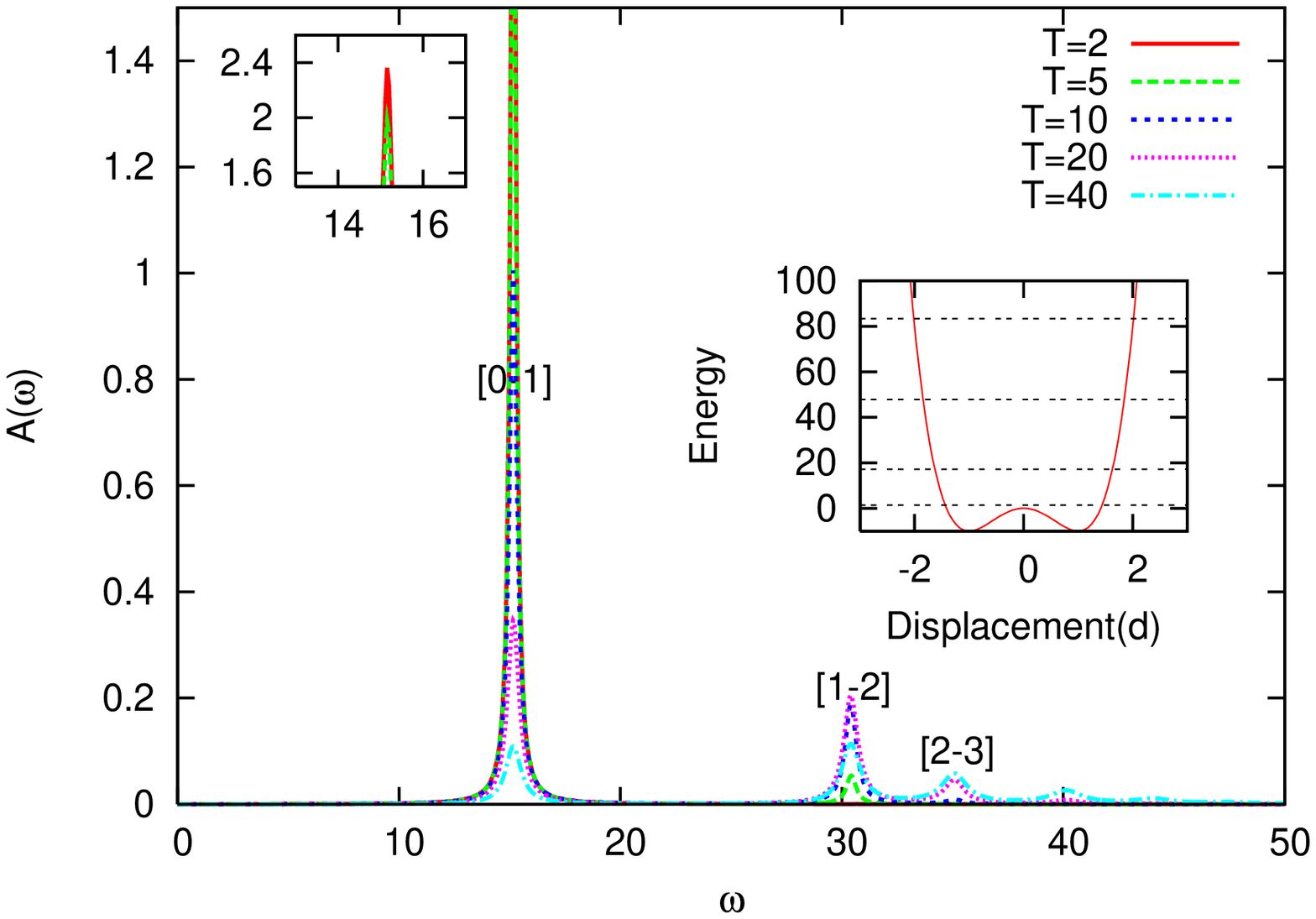}
   \caption{(a,b)=(-2,1)$\hspace{3mm}$The notations used in the figure are the same as Fig.\ref{0.98_0.02}}
   \label{m2_1}
\end{center}
\end{figure}
\begin{figure}[h]
\begin{center}
  \includegraphics[width=0.7\hsize,angle=-90]{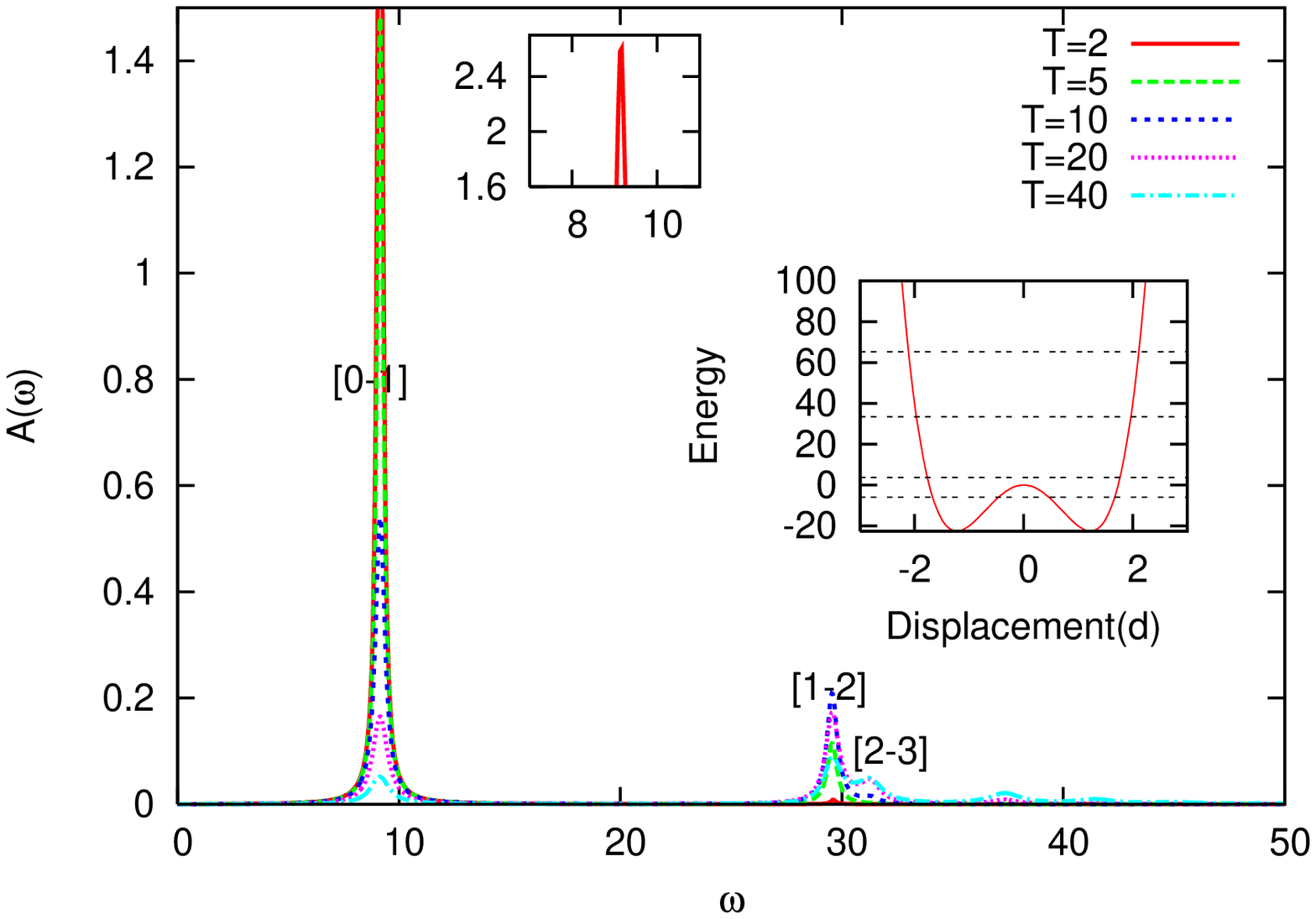}
   \caption{(a,b)=(-3,1)$\hspace{3mm}$The notations used in the figure are the same as Fig.\ref{0.98_0.02}}
   \label{m3_1}
\end{center}
\end{figure}
\begin{figure}[h]
\begin{center}
  \includegraphics[width=0.7\hsize,angle=-90]{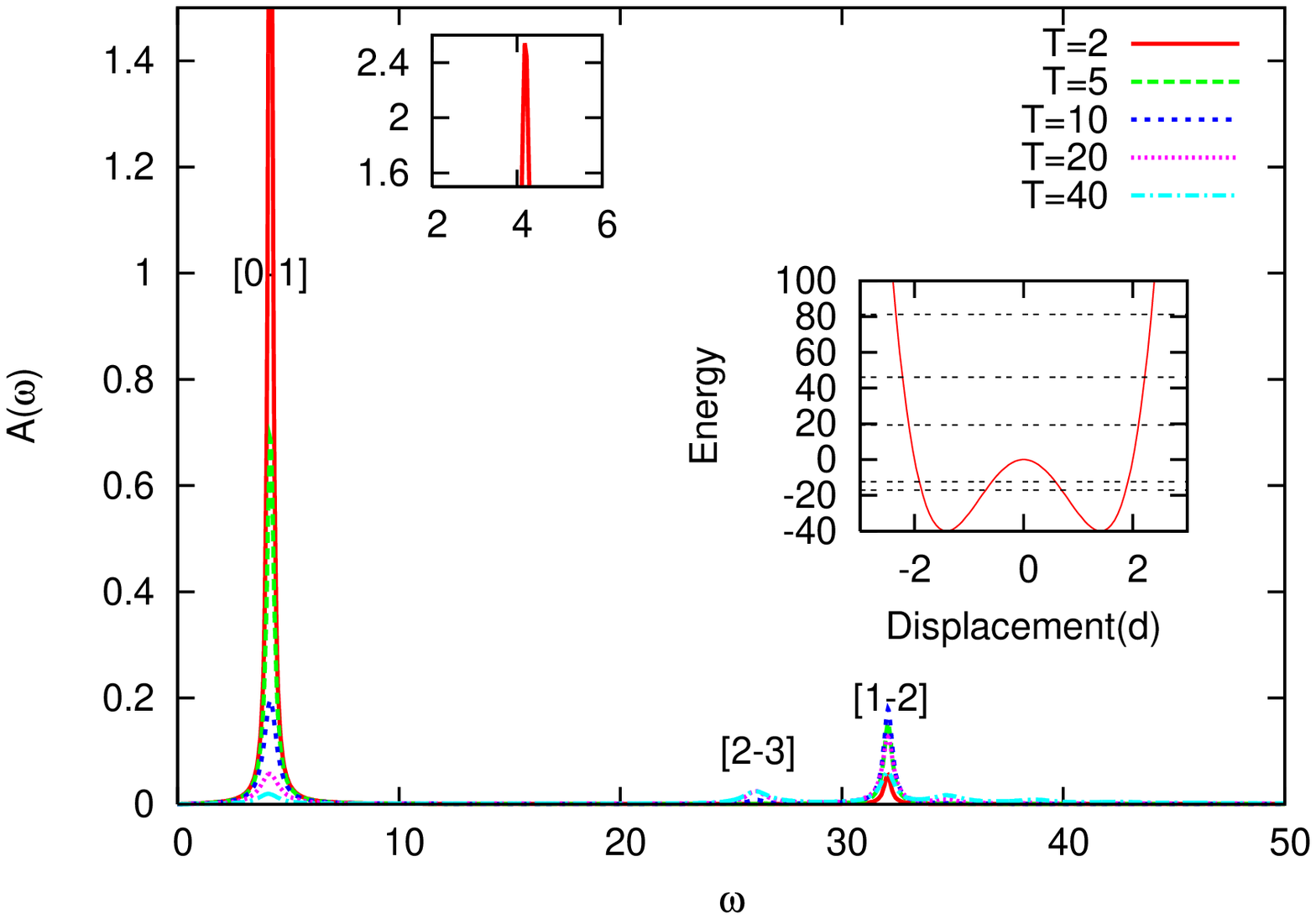}
   \caption{(a,b)=(-4,1)$\hspace{3mm}$The notations used in the figure are the same as Fig.\ref{0.98_0.02}}
   \label{m4_1}
\end{center}
\end{figure}
Close to the harmonic limit like Fig.\ref{0.98_0.02}, almost degenerate peaks sit and the spectral weight shifts to higher energies
with increasing temperature. This shift is qualitatively consistent with the
description of the self-consistent quasiharmonic approximation\cite{DU}. In this limit, many peaks are united and it appears that the damping is enhanced effectively. 
As anharmonicity increases(Fig.\ref{0.9_0.1}$\rightarrow$\ref{0_1}), 
excitation energy to the neighboring state becomes large progressively as the energy of the initial state increases. Then, we observe the sequence of peaks 
of which the weight shifts to higher energy monotonously with increasing temperature. As the coefficient of the square
term of ${\tilde x}$ become large to
negative value(Fig.\ref{m1_1}$\rightarrow$\ref{m4_1}), quasi degeneracy between the bonding and anti-bonding
states appears. In this range, we can divide transitions into two groups, one of which expresses the transition between quasi-degenerated states
lying in lower energy range, and the other expresses transitions between non-degenerated states lying in higher energy range.
An interesting case is the shoulder at $\omega \sim 30$ seen in 
Fig.\ref{m3_1}. This characteristic spectral shape is due to
the crossover of the above mentioned two sequences. Between $a=-3$ and $a=-4$,
the transition energy denoted by [2,3] gets smaller than that of [1,2],
then nonmonotonous tendency of the peak height which corresponds to the Boltzmann weight of the initial states appears.
In the off-center limit
($a\rightarrow-\infty$), the ion eigenstates approaches the linear combination of two off-center oscillations.
Peaks will be consolidated to two peaks of which the lower one approaches zero energy and the other denotes the
energy of the off-center oscillation. 

\begin{figure}
 \begin{center}
    \includegraphics[width=0.7\hsize, angle=-90]{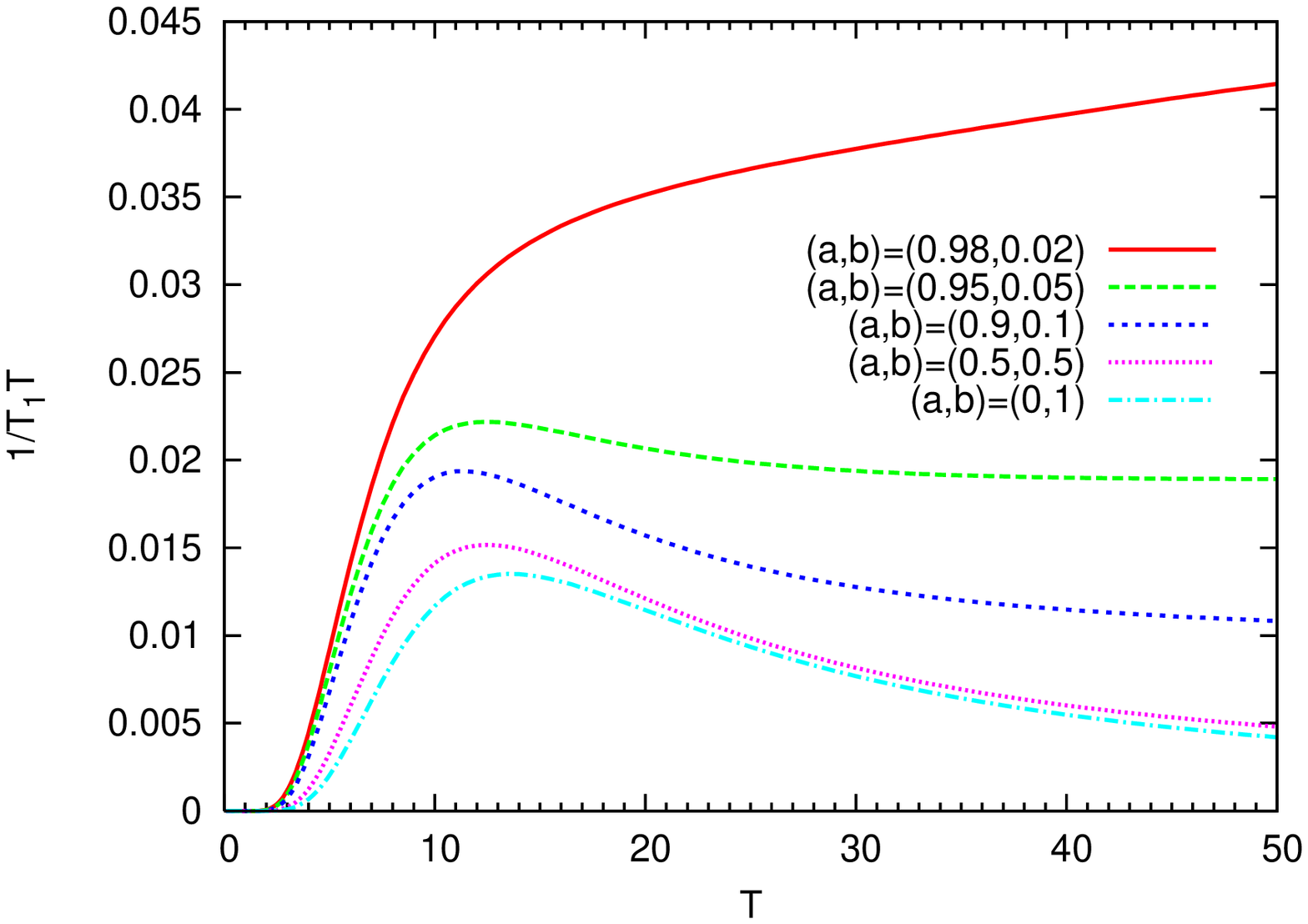}
    \caption{NMR relaxation rate due to the two phonon Raman process at the ion site(Eq.(\ref{local})) with $a\geq 0$.}
     \label{nmrp}
 \end{center}
\end{figure}

\begin{figure}
 \begin{center}
    \includegraphics[width=0.7\hsize, angle=-90]{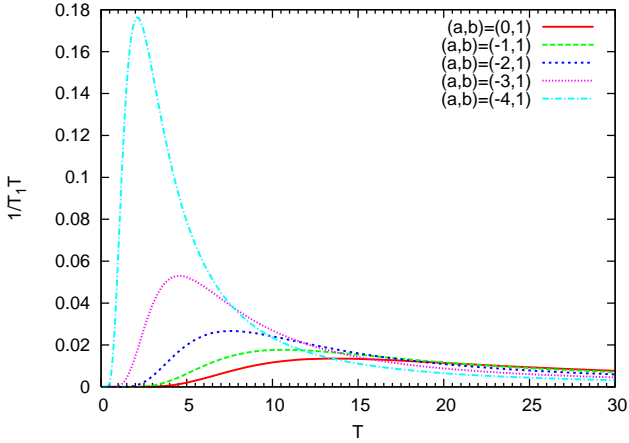}
     \caption{NMR relaxation rate due to the two phonon Raman process at the ion site(Eq.(\ref{local})) with $a\leq 0$.}
     \label{nmrm}
 \end{center}
\end{figure}

NMR relaxation rate due to the two-phonon Raman process is expressed as follows\cite{DU}.
\begin{equation}
\frac{1}{T_1^R}\propto \int _{-\infty}^{\infty} \!\!\!\!\ {\rm d}\omega \hspace{1mm} A^2(\omega)\left[n(\omega)+1\right]n(\omega)
\end{equation}
where $n(\omega)$ denotes the Bose distribution function. The plot of $1/T_1^RT$ for various type of potential is shown in Fig.\ref{nmrp} and \ref{nmrm}.
In the harmonic limit like $(a,b)=(0.98,0.02)$, $1/T_1^RT$ increases monotonously. Then, with increasing anharmonicity, we see that the low-energy peak
which is observed in the experiment\cite{nmr1} develops. That change is consistent with the result of the self-consistent
quasi-harmonic approximation\cite{DU}.

The behavior of $1/T_1^RT$ with increasing anharmonicity is explained in relation to the spectral function as
follows. If anharmonicity is strong enough, 
the spectral weight shift to higher energy takes place faster than the gain of Bose function weight.
Then low energy peak appears as shown in Fig.\ref{nmrp}.
As the low frequency peak in the spectrum grows towards the off-center limit, the peak in the 
NMR relaxation rate gets sharpened as shown in Fig.\ref{nmrm}. 

\section{Summary and Discussion}
In this paper, we have derived the general expression of the ion Green's function in the second order perturbation of electron-ion coupling.
Explicit forms of the self-energy, vertex-correction, and normalization factor are obtained
independent from the strength of the anharmonicity. Although we treated the oscillator moving in one dimension, it is
straightforward to generalize the results to a three dimensional oscillator. In this sense, the present theoretical
scheme may be useful for a wide class of problems concerned with anharmonic lattice vibration. 

Using the expression, we have discussed characteristic properties of ion spectral functions in the entire range
of anharmonicity, from almost harmonic case to the well-separated
double-well potential. These spectrums are observable by THz absorption
and we expect that experimental results\cite{Mori} can be understood qualitatively 
as the combination of characteristics discussed in the previous section for some typical cases. 
Additionally, we calculated NMR relaxation rate due to the two phonon Raman process, and reproduced the
low temperature peak. The spectrum of the anharmonic lattice vibration is naturally reflected in other experiments. The present work
will provide a basis of analysis of ultrasonic measurements\cite{Goto2004-1,Goto2004-2,Goto2005} and Raman scattering experiments\cite{Takasu2006,Hasegawa2008,Takasu2008}
on $\beta$-pyrochlore, skutterudite, and clathrate compounds.

The higher order perturbation is possible at least in principle using continued fractions, but it will be
cumbersome and laborious task.
About the electronic properties, it may be straightforward to apply similar procedure based on the second order
perturbation expansion. The merit of the method in this paper is that it allows us  to
consider full effect of anharmonicity as we have stressed already. In the weak coupling region, it may be possible to construct a microscopic theory which will reveal the
relation between anharmonic lattice vibration
and electron mass-enhancement, and furthermore superconductivity eventually.  
However, we will leave these interesting problems as future study. 
\section*{Acknowledgments}
The authors would like to thank Thomas Dahm for useful discussions. They are also grateful to Prof. N. Toyota for informing their
experimental data prior to publication.
This work is supported by a Grant-in-Aid for Scientific Research (C) (No.20540347) from Japan Society for the Promotion of Science.
\appendix
\section{Explicit expressions of Self-energy, vertex correction, and normalization factor}
We obtain the expression of self-energy, vertex correction, and normalization factor as
Eq.(\ref{AP_selfenergy}),(\ref{AP_vertexcorrection}),(\ref{AP_normalizationfactor})
by following the procedure written in the Sec.\ref{sec:main}.
\begin{table*}
\begin{eqnarray}
&&\Pi({\rm i}\omega_n,i_1,i_4)= 2 \sum_{{\bf kq}}|V({\bf q})|^2 f(\xi_{{\bf k}+{\bf q}})(1-f(\xi_{{\bf k}}))
  \sum_{i_2} \frac{|\langle i_1|x|i_2 \rangle|^2}{{\rm e}^{-\beta E_{i_1}}-{\rm e}^{-\beta E_{i_4}}}\times \nonumber \\
&&\hspace{-10mm}\Bigl[-{\rm e}^{-\beta E_{i_4}}({\rm i}\omega_n+E_{i_2}-E_{i_4}+\xi_{{\bf k}}-\xi_{{\bf k}+{\bf q}})^{-1}-{\rm e}^{-\beta E_{i_2}}(-{\rm i}\omega_n-E_{i_2}+E_{i_4}+\xi_{{\bf k}}-\xi_{{\bf k}+{\bf q}})^{-1} \nonumber \\
&&      +{\rm e}^{-\beta E_{i_2}}(E_{i_1}-E_{i_2}+\xi_{{\bf k}}-\xi_{{\bf k}+{\bf q}})^{-1}       +{\rm e}^{-\beta E_{i_1}}(E_{i_2}-E_{i_1}+\xi_{{\bf k}}-\xi_{{\bf k}+{\bf q}})^{-1} \Bigr]\nonumber \\
\nonumber \\
&& +2 \sum_{{\bf kq}}|V({\bf q})|^2 f(\xi_{{\bf k}+{\bf q}})(1-f(\xi_{{\bf k}}))
   \sum_{i_3} \frac{|\langle i_3|x|i_4 \rangle|^2}{{\rm e}^{-\beta E_{i_1}}-{\rm e}^{-\beta E_{i_4}}}\times \nonumber \\
&&\hspace{-10mm}\Bigl[-{\rm e}^{-\beta E_{i_1}}(-{\rm i}\omega_n-E_{i_1}+E_{i_3}+\xi_{{\bf k}}-\xi_{{\bf k}+{\bf q}})^{-1}-{\rm e}^{-\beta E_{i_3}}({\rm i}\omega_n+E_{i_1}-E_{i_3}+\xi_{{\bf k}}-\xi_{{\bf k}+{\bf q}})^{-1} \nonumber \\
&&    +{\rm e}^{-\beta E_{i_3}}(E_{i_4}-E_{i_3}+\xi_{{\bf k}}-\xi_{{\bf k}+{\bf q}})^{-1}        +{\rm e}^{-\beta E_{i_4}}(E_{i_3}-E_{i_4}+\xi_{{\bf k}}-\xi_{{\bf k}+{\bf q}})^{-1} \Bigr]\label{AP_selfenergy}
\end{eqnarray}
\end{table*}
\begin{table*}
\begin{eqnarray}
&&\delta x({\rm i}\omega_n,i_1,i_4)= 2 \sum_{{\bf kq}}|V({\bf q})|^2 f(\xi_{{\bf k}+{\bf q}})(1-f(\xi_{{\bf k}}))
  \sum_{i_2  i_3}\langle i_1|x|i_2 \rangle\langle i_2|x|i_3 \rangle\langle i_3|x|i_4 \rangle\langle i_4|x|i_1 \rangle \times \nonumber \\
&&\Bigl[2\bigl\{(1-\delta_{i_1 i_3})(E_{i_3}-E_{i_1})^{-1}-(E_{i_4}-E_{i_3}+E_{i_2}-E_{i_1})^{-1}\bigr\}\frac{{\rm e}^{-\beta E_{i_4}}}{{\rm e}^{-\beta E_{i_1}}-{\rm e}^{-\beta E_{i_4}}} \nonumber \\
&&\hspace{10mm}\times\bigl\{(E_{i_2}-E_{i_1}+\xi_{{\bf k}}-\xi_{{\bf k}+{\bf q}})^{-1}-({\rm i}\omega_n+E_{i_2}-E_{i_4}+\xi_{{\bf k}}-\xi_{{\bf k}+{\bf q}})^{-1}\bigr\}\nonumber\\
&&+2\bigl\{(1-\delta_{i_2 i_4})(E_{i_4}-E_{i_2})^{-1}-(E_{i_4}-E_{i_3}+E_{i_2}-E_{i_1})^{-1}\bigr\}\frac{{\rm e}^{-\beta E_{i_3}}}{{\rm e}^{-\beta E_{i_1}}-{\rm e}^{-\beta E_{i_4}}} \nonumber \\
&&\hspace{10mm}\times\bigl\{(E_{i_4}-E_{i_3}+\xi_{{\bf k}}-\xi_{{\bf k}+{\bf q}})^{-1}-({\rm i}\omega_n+E_{i_1}-E_{i_3}+\xi_{{\bf k}}-\xi_{{\bf k}+{\bf q}})^{-1}\bigr\}\nonumber\\
&&+2\bigl\{(1-\delta_{i_1 i_3})(E_{i_3}-E_{i_1})^{-1}-(E_{i_4}-E_{i_3}+E_{i_2}-E_{i_1})^{-1}\bigr\}\frac{{\rm e}^{-\beta E_{i_2}}}{{\rm e}^{-\beta E_{i_1}}-{\rm e}^{-\beta E_{i_4}}} \nonumber \\
&&\hspace{10mm}\times\bigl\{(E_{i_1}-E_{i_2}+\xi_{{\bf k}}-\xi_{{\bf k}+{\bf q}})^{-1}-(-{\rm i}\omega_n-E_{i_2}+E_{i_4}+\xi_{{\bf k}}-\xi_{{\bf k}+{\bf q}})^{-1}\bigr\}\nonumber\\
&&+2\bigl\{(1-\delta_{i_2 i_4})(E_{i_4}-E_{i_2})^{-1}-(E_{i_4}-E_{i_3}+E_{i_2}-E_{i_1})^{-1}\bigr\}\frac{{\rm e}^{-\beta E_{i_1}}}{{\rm e}^{-\beta E_{i_1}}-{\rm e}^{-\beta E_{i_4}}} \nonumber \\
&&\hspace{10mm}\times\bigl\{(E_{i_3}-E_{i_4}+\xi_{{\bf k}}-\xi_{{\bf k}+{\bf q}})^{-1}-(-{\rm i}\omega_n-E_{i_1}-E_{i_3}+\xi_{{\bf k}}-\xi_{{\bf k}+{\bf q}})^{-1}\bigr\}\nonumber\\
&&+(1-\delta_{i_1 i_3})(E_{i_3}-E_{i_1})^{-1}\bigl\{(E_{i_2}-E_{i_1}+\xi_{{\bf k}}-\xi_{{\bf k}+{\bf q}})^{-1}+(E_{i_2}-E_{i_3}+\xi_{{\bf k}}-\xi_{{\bf k}+{\bf q}})^{-1}\bigr\}\nonumber\\
&&-(1-\delta_{i_2 i_4})(E_{i_4}-E_{i_2})^{-1}\bigl\{(E_{i_3}-E_{i_4}+\xi_{{\bf k}}-\xi_{{\bf k}+{\bf q}})^{-1}+(E_{i_3}-E_{i_2}+\xi_{{\bf k}}-\xi_{{\bf k}+{\bf q}})^{-1}\bigr\}\nonumber\\
&&-2(E_{i_4}-E_{i_3}+E_{i_2}-E_{i_1})^{-1}\nonumber\\
&&\hspace{10mm}\times\bigl\{(E_{i_2}-E_{i_1}+\xi_{{\bf k}}-\xi_{{\bf k}+{\bf q}})^{-1}-(E_{i_3}-E_{i_4}+\xi_{{\bf k}}-\xi_{{\bf k}+{\bf q}})^{-1}\bigr\}\Bigr]\label{AP_vertexcorrection}
\end{eqnarray}
\end{table*}
\begin{table*}
\begin{eqnarray}
&&\Lambda(i_1,i_4)=2\beta \sum_{{\bf kq}}|V({\bf q})|^2 \frac{f(\xi_{{\bf k}+{\bf q}})(1-f(\xi_{{\bf k}}))}{{\rm e}^{-\beta E_{i_1}}-{\rm e}^{-\beta E_{i_4}}}\times \nonumber \\
&&\hspace{-1cm}\Bigl[{\rm e}^{-\beta E_{i_1}}\sum_{i_2} |\langle i_1|x|i_2 \rangle|^2(E_{i_2}-E_{i_1}+\xi_{{\bf k}}-\xi_{{\bf k}+{\bf q}})^{-1} -
 {\rm e}^{-\beta E_{i_4}}\sum_{i_3} |\langle i_3|x|i_4 \rangle|^2(E_{i_3}-E_{i_4}+\xi_{{\bf k}}-\xi_{{\bf k}+{\bf q}})^{-1}\Bigr] \nonumber\\
\nonumber \\
&&-2\beta Z_{i0}^{-1} \sum_{{\bf kq}}|V({\bf q})|^2 f(\xi_{{\bf k}+{\bf q}})(1-f(\xi_{{\bf k}})) \sum_{i_2i_3}  {\rm e}^{-\beta E_{i_2}}|\langle i_2|x|i_3 \rangle|^2 
(E_{i_3}-E_{i_2}+\xi_{{\bf k}}-\xi_{{\bf k}+{\bf q}})^{-1}\label{f20}\nonumber\\
&&+2 \sum_{{\bf kq}}|V({\bf q})|^2\sum_{i_2} \frac{|\langle i_1|x|i_2 \rangle|^2}{{\rm e}^{-\beta E_{i_1}}-{\rm e}^{-\beta E_{i_4}}}
(E_{i_2}-E_{i_1}+\xi_{{\bf k}}-\xi_{{\bf k}+{\bf q}})^{-2} \times \nonumber \\
&&\hspace{2cm}\left[{\rm e}^{-\beta E_{i_2}}f(\xi_{{\bf k}})(1-f(\xi_{{\bf k}+{\bf q}}))-{\rm e}^{-\beta E_{i_1}}f(\xi_{{\bf k}+{\bf q}})(1-f(\xi_{{\bf k}}))\right]  \nonumber\\
\nonumber \\
&&-2 \sum_{{\bf kq}}|V({\bf q})|^2\sum_{i_3} \frac{|\langle i_3|x|i_4 \rangle|^2}{{\rm e}^{-\beta E_{i_1}}-{\rm e}^{-\beta E_{i_4}}}
(E_{i_3}-E_{i_4}+\xi_{{\bf k}}-\xi_{{\bf k}+{\bf q}})^{-2} \times \nonumber \\
&&\hspace{2cm}\left[{\rm e}^{-\beta E_{i_3}}f(\xi_{{\bf k}})(1-f(\xi_{{\bf k}+{\bf q}}))-{\rm e}^{-\beta E_{i_4}}f(\xi_{{\bf k}+{\bf q}})(1-f(\xi_{{\bf k}})) \right] \label{AP_normalizationfactor}
\end{eqnarray}
\end{table*}

\end{document}